\documentclass[10pt,a4paper]{article}
\usepackage[pdftex]{graphicx} 
\usepackage[utf8]{inputenc}
\usepackage{amsmath}
\usepackage{amsfonts}
\usepackage{amssymb}
\usepackage{subfig}
\usepackage{color}
\usepackage{mathptmx}
\usepackage[toc,title]{appendix}

\newcommand{\etal}{\textit{et al.}}

   \graphicspath{{graph/}}
\usepackage[left=2cm,right=2cm,top=2cm,bottom=2cm]{geometry}
\author{Srijita Sinha\footnote{Email: ss13ip012@iiserkol.ac.in}\hspace{0.15cm} and Narayan Banerjee\footnote{Email: narayan@iiserkol.ac.in} \\  IISER Kolkata, Mohanpur Campus, Mohanpur, Nadia 741246, India}
\title{Density perturbation in an interacting holographic dark energy model}
\begin{document}
\date{}
\maketitle
\begin{abstract}
The present work deals with the evolution of the density contrasts for a cosmological model where along with the standard cold dark matter (CDM), the present Universe also contains holographic dark energy (HDE). The HDE is allowed to interact with the CDM. The equations for the density contrasts are integrated numerically. It is found that irrespective of the presence of an interaction, the matter perturbation has growing modes. The HDE is also found to have growth of perturbation, so very much like the CDM, HDE can also cluster. The interesting point to note is that the density contrast corresponding to HDE has a peak at a recent past and is decaying at the present epoch.
\end{abstract}

\vspace{0.5cm}
\vspace{0.5 cm}
\section{Introduction}

Ever since the discovery through the luminosity versus redshift surveys\cite{perlmutter1999aj, schmidt, riess1998aj}, that the Universe at the present epoch is expanding with an acceleration, there has been a proliferation of proposals of a ``dark energy'' that gravitates in the wrong way. A cosmological constant $\Lambda$ appears to be a very competent candidate although not a clear winner because of the insurmountable discrepancy between the observational requirement and the theoretically predicted value. A scalar field with a suitable potential, called a quintessence, is arguably the close second. There are excellent reviews that summarise the list of candidates and their strength and weakness \cite{paddy2003pr, carroll, sahni2000ijmpd, copeland2006ijmpd}. Among the quintessence models, some may evolve in a way such that the equation of state parameter of the dark energy attains a value less that $-1$ at the present epoch or in a finite future\cite{feng2005plb}. Such models are called ``quintom'' models as they evolve to mimic a phantom model. Several representative models of quintom cosmology can also be found in \cite{cai2007plb1,  cai2007plb2, cai2008cqg}. Moreover, this dark energy model can yield important cosmological applications, namely, it may avoid the big bang singularity if applied to the very early Universe. This result can be found in \cite{cai2007jhep} and \cite{ cai2012jcap}. For a more detail theoretical and observational implications we refer to the work of Cai \etal\cite{cai2010pr}.

Another most talked about form of dark energy is the so-called holographic dark energy (HDE) based on the holographic principle in quantum gravity theory\cite{susskind}. The holographic principle, following the 't Hooft conjecture\cite{thooft}, that the information contained in a volume can be ascertained with the knowledge about the degrees of freedom residing on its boundary actually stems from Bekenstein's idea that the entropy of a black hole is related to its area\cite{beken7}. Based on the holographic principle, one of the characteristic features of the HDE is the long distance cut-off, called the infra-red (IR) cut-off\cite{cohen}. In the context of cosmology, this cut-off is not uniquely specified but rather realised in various ways. One of the natural choices is the Hubble radius\cite{xu}. The drawback of this model, as shown by Hsu\cite{hsu2004}, is that it does not provide the recent acceleration. In this context, Zimdahl and Pav\'{o}n in \cite{diego,zimdahl} showed that allowing a non-gravitational interaction in the dark sector of the Universe not only solves this problem but also alleviate the nagging coincidence problem\cite{diego1, hu}. Other possibilities are the particle horizon as suggested by Fischler and Susskind\cite{fischler1998} and Bousso\cite{bousso1999} and the future event horizon as suggested by Li\cite{lim} and Huang and Li\cite{huang2004}. A more recent choice for the cut-off scale is the Ricci scalar curvature used by Gao \etal\cite{gao}, Feng\cite{feng2008, feng231}, to name a few. Some of the notable work in modified Ricci Holographic DE include \cite{chimento2011mpj, chimento2011,chimento2012,chimento2013epjc1,chimento2013epjc2,nojiri2017}. A generalized holographic inflation as well as  holographic bounce with no specific IR cut-off has also been proposed\cite{nojiri2019plb,nojiri2019npb}.

Ho\v{r}ava and Minica showed that the expectation value of the cosmological constant is zero in the context of holographic principle \cite{horava}. This feature made holographic dark energy quite an attractive candidate as the dark energy. There had been an attempt to find a unified model giving an early inflation and a late time acceleration by Nojiri and Odintsov\cite{odint}. Holographic dark energy models have also been considered in some modified theories of gravity, for example, in Brans-Dicke theory by Setare \cite{setare}, Banerjee and Pav\'{o}n \cite{nb1}, in Gauss-Bonnet gravity by Saridakis \cite{saridakis}, Setare \cite{setare1}. As there is a host of observational data in cosmology, almost all kinds of holographic dark energy models are now tested against observations. Some of them are by Campo \etal\cite{campo}, Zhang and Wu\cite{wu}, Li \etal\cite{lim2}, Feng \etal\cite{su}, Mukherjee \etal\cite{ankan} and a recent one by D’Agostino\cite{dagostino2019}. The list is far from being exhaustive. The interacting and non-interacting HDE models have been studied in detail using the Planck data by Li \etal \cite{yun2013}, Li \etal\cite{li2013jcap}, Zhang \etal\cite{zhang2015} and Feng \etal \cite{feng2018}. The stability criteria for holographic dark energy have also been discussed widely\cite{li2008, nb2, mahata}. Although the HDE model is quite lucrative in many ways, it cannot avoid the phantom Universe\cite{xin2005,xu2009}. One way of preventing the future ``big-rip'' singularity is to allow some phenomenological interaction between the dark matter and dark energy as shown by Wang \etal \cite{wang141, wang357}. Thereafter the interacting HDE model has been studied extensively in\cite{mohseni231,sen-diego, zhang26, wang1, diego-sen} and references therein.

The HDE, as it produces the recent acceleration of the Universe, will also affect the formation of large scale structures. Moreover, the evolution of the matter density perturbation can provide some knowledge about the components affecting it, in this case, the HDE. Since HDE is an evolving component of the Universe, it will have fluctuations like that of the DM and hence will not only affect the growth of matter perturbations\cite{abramo} but also may cluster on its own\cite{mehrabi1478, batista2013}. There are some work in the matter perturbation in HDE models, such as the one by Kim \etal\cite{kim} for a decaying HDE. Very recently, perturbation in a clustered HDE has been investigated by Mehrabi \etal\cite{mehrabi} and Malekjani \etal\cite{malekjani2018}. 

In this work, we will show how the evolution of DM, as well as DE, are affected in the presence of an interaction between them. In fact, this is the first full relativistic treatment of the perturbation of holographic dark energy models to start with. The closest work to this is the one by Mehrabi \etal\cite{mehrabi}, where a relativistic treatment is definitely given, but the evolution equation of the gravitational potential $\Phi$ is approximated to the standard Poisson's equation. Another important new feature is certainly the inclusion of interaction between the HDE and DM in the perturbation. The inclusion of interaction leads to a brief transient oscillatory period for the density contrast for both HDE and DM. This existence of growing mode with a transient oscillatory behaviour is an entirely new feature. A peak in the dark energy density contrast in the recent past (low values of $z$) is another new feature which escaped the notice so far.

The paper is organised as follows. Section \ref{sec:backgrnd} discusses the background, section \ref{sec:int-hde} deals with a scenario where an interaction between the cold dark matter and the holographic dark energy is allowed. Section \ref{sec:pert} includes the relevant equations that describe the perturbation in the present case. The evolution of the density contrast is discussed in section \ref{sec:evolution}. In the last section, section \ref{sec:summary}, we summarise and discuss the results. We also include an appendix, where the coefficients of the second order differential equations are explicitly written, which are not done in the main text in order to improve the readability.

\section{The background} \label{sec:backgrnd}

We consider a spatially flat, homogeneous and isotropic Universe, given by the Friedman-Robertson-Walker (FRW) metric,
\begin{equation}\label{metric}
ds^2= a^2(\eta)\left(- d \eta ^2+\gamma_{ij} d x^i dx^j\right),
\end{equation}
where $\gamma^i_{j}=\delta^i_j $. The Friedmann equations for this metric take the form
\begin{eqnarray}
3 \mathcal{H}^2 &=& -a^2 \kappa \rho ,\\ \label{fd1}
\mathcal{H}^2+ 2 \mathcal{H}^\prime &=& a^2 \kappa p , \label{fd2}
\end{eqnarray}
where $a(\eta)$ is the scale factor, $\kappa=8 \pi G_N$ ($G_N$ being the Newtonian Gravitational constant), $\rho(\eta)$ and $p(\eta)$ are the total density and the total pressure of the matter distribution in the Universe respectively, $\mathcal{H}= \frac{a^\prime}{a}$ is the Hubble parameter in conformal time $\eta$ and prime $(^\prime)$ denotes the derivative with respect to the conformal time. The conformal time $\eta$ is related to the cosmic time $t$ as $a^2 d\eta^2 = dt^2$. Hereafter, the Greek indices $\mu , \nu \dotsc $ denote the space-time coordinates while the Latin indices $i ,j \dotsc$ denote the coordinates in the spatial hypersurface.

It is assumed that the Universe is filled with a perfect fluid dominated by a pressureless (cold) dark matter (CDM) and holographic dark energy (HDE). The energy densities and pressure are such that  $\rho=\rho_m+\rho_{de}$ and $p=p_{de}$ respectively. Subscript `\textit{m}' denotes the contribution of the CDM while `\textit{de}' denotes that of the HDE. There is an interaction between the two components of the Universe, CDM and HDE, hence a transfer of energy between the two.  The total energy balance equation 
\begin{equation}
 \rho^\prime + 3 \mathcal{H} (\rho + p) = 0 \label{cons},
\end{equation}
is thus divided into two equations, 
\begin{eqnarray}
\rho^\prime_m+ 3 \mathcal{H} \rho_m &=& aQ ,\label{cons1}\\
\rho^\prime_{de}+ 3 \mathcal{H} \left(1+w_{de}\right) \rho_{de}&=& -aQ \label{cons2},
\end{eqnarray}
where $Q$ is the rate of energy density transfer, $w_{de}= \frac{p_{de}}{\rho_{de}}$ is the equation of state (EoS) parameter for the HDE. It is clear that equations (\ref{cons1}) and (\ref{cons2}) together give the conservation equation. The non-interacting scenario can be recovered simply by setting $Q=0$. If $Q>0$ energy is transferred from dark energy to dark matter and vice versa. There is, however, no compelling observational binding to take this interaction into account, but as the two sectors, DM and DE evolve together, this interaction adds to the generality of the model. There are quite a few investigations regarding this interaction. For a review, we refer to the work by Wang {\it et al}\cite{wang2016}. Some very recent work indicate that the observational anomaly of the 21-cm line excess at cosmic dawn can be the relevant observations in this connection\cite{li2019plb, li2020plb}.

The expression for the energy density of HDE is
\begin{equation}
\rho_{de}=3 C^2 M_{P}^2 L^2, \label{def}
\end{equation}
where $3C^2$ is a numerical constant introduced for convenience, $M_{P}=\sqrt{\frac{1}{\kappa}}$ is the reduced Plank mass, $L$ is the characteristic length scale of the Universe which provides the IR cut-off of $\rho_{de}$. Incidentally, in the present work this cut-off is chosen as the future event horizon as suggested by Li\cite{lim}, 
\begin{equation}
L = a \int_t ^\infty\frac{d \tilde{t}}{a} = a \int_a^\infty \frac{d \tilde{a}}{H \tilde{a}^2} ,
\end{equation}
$H$ is the Hubble parameter in cosmic time $t$. It has already been mentioned in the introduction that this is by no means the only choice or the best choice as the infra-red cut-off. 

The energy-momentum tensor of the fluid `A' (which stands for either `\textit{m}' or `\textit{de}') is $T^\mu_{\left(A\right) \nu}$ and is given by 
\begin{equation} \label{stressA}
T^\mu_{\left(A\right) \nu}= \left(\rho_A + p_A \right)u^\mu_{\left(A\right)} u_{\left(A\right)\nu} +  p_A  \delta^\mu{}_\nu \hspace{0.5mm} ,
\end{equation}
where $u_{\left(A\right)\mu}= -a \delta^0_\mu$ is the comoving $4$-velocity of the fluid `A'. The total energy-momentum tensor is $T^\mu{}_\nu= \sum_A T^\mu_{\left(A\right)\nu}$ such that
\begin{equation} \label{stress}
\begin{split}
\left(\rho+p\right)u{}^\mu u_\nu +p \delta^\mu{}_\nu = \sum_A \left(\rho_A + p_A \right)u^\mu_{\left(A\right)} u_{\left(A\right)\nu} + \sum_A p_A  \delta^\mu{}_\nu,
\end{split} 
\end{equation}
where  $u_\mu = -a \delta^0_\mu$ is the total comoving $4$-velocity . It is again clear from equation (\ref{stress}) that $\rho=\sum_A \rho_A$, $p=\sum_A p_A$.

In presence of an interaction, the energy-momentum tensor of the individual components $T^\mu_{\left(A\right) \nu}$ does not conserve independently, and its divergence has the source term  $Q_{\left(A\right)\nu}$. Thus the covariant form of the conservation equation for fluid `A' is given as
\begin{equation} \label{condition}
T^\mu_{\left(A\right) \nu; \mu} =Q_{\left(A\right)\nu} ~, \hspace{0.2cm} \mbox{where}  \hspace{0.2cm} \sum_A Q_{\left(A\right)\nu} =0 ~.
\end{equation}
The source term for the interaction is a $4$-vector and has the form
\begin{equation} \label{Q-def}
Q^\mu_m=\frac{1}{a} \left(Q_m, \vec{0}\right) = \frac{1}{a} \left(Q, \vec{0}\right)= \frac{1}{a} \left(-Q_{de}, \vec{0}\right) = -Q^\mu_{de}~.
\end{equation}
It is assumed that there is no momentum transfer in the background universe. The energy balance equation for the fluid `A'  takes the form
\begin{equation}\label{balaneA}
\rho'_{A}+ 3 \mathcal{H} \left(1+w_{A}\right) \rho_{A}= a Q_A ~,
\end{equation}
where $Q_A= Q_{(A)0}$, the time component of the four vector $Q_{(A)\mu}$ .

\section{Interacting Holographic Dark Energy} \label{sec:int-hde}

The evolution of the dimensionless HDE density parameter $\Omega_{de}=\frac{\rho_{de}}{\rho_c}$ where $\rho_c= 3 H_0^2 M_{P}^2$ is the critical density of the Universe, and the dimensionless Hubble parameter $E$, in the presence of an interaction are governed by the simultaneous differential equations \cite{zhang12}
\begin{equation} \label{int-omega}
\frac{d \Omega_{de}}{d z} =-\frac{2\Omega_{de}\left( 1 -\Omega_{de} \right)}{1+z} \left( \frac{1}{2}+ \sqrt{\frac{\Omega_{de}}{C^2}}-\frac{\Omega_I}{2\left(1-\Omega_{de}\right)}\right),
\end{equation}
\begin{equation} \label{int-hubble}
\frac{1}{E}\frac{d E}{d z} = -\frac{ \Omega_{de}}{1+z} \left( \frac{1}{2}+\sqrt{\frac{\Omega_{de}}{C^2}} +\frac{\Omega_I- 3}{2 \Omega_{de}} \right),
\end{equation}
where $E =\frac{H}{H_0}$ is the Hubble parameter scaled by its present value $H_0$. The evolution equations 
(\ref{int-omega}) and (\ref{int-hubble}) are given in terms of the cosmic redshift $z$, which is a dimensionless quantity and is related to the scale factor $a$ as $z = \frac{a_0}{a} -1$, $a_0$ being the present value of the scale factor (taken to be unity). These two equations are obtained following the usual steps (also shown in \cite{zhang12}).

The EoS parameter of DE, $w_{de}$, is an intrinsic characteristic of DE. From the system of equations, given by Einstein's equations and the conservation equations, $w_{de}$ imposes a constraint on $Q$ (see \cite{zhang12}) as
\begin{equation} \label{int-eos}
w_{de}=-\frac{1}{3}-\frac{2}{3}\sqrt{\frac{\Omega_{de}}{C^2}}-\frac{\Omega_I}{3 \Omega_{de}}~,
\end{equation}
where $\Omega_I=\frac{Q}{H  \rho_c}$ is the interaction term expressed in a dimensionless form. 
 
To study the effect of interaction, we need to take a specific form of the interaction term $Q$. Models with interaction term $Q$ proportional to either $\rho_m$ or $\rho_{de}$ or any combination of them have been studied extensively in literature\cite{ zhang12, valiviita2008jcap, feng, miaoli09, li2, valiviita2015jcap, wang2016, acosta, funo, clemson2012prd, bohmer}. It should be noted that there is no theoretical or observational compulsion for any one of these choices. In the present work we consider $Q \propto \rho_{de}$. We have taken the covariant form of the source term $Q^{\mu}_{m}\left(\eta\right) $ as
\begin{equation} \label{interaction}
Q^{\mu}_{m} = - Q^{\mu}_{de} =  \frac{\beta \mathcal{H} \rho_{de} u^{\mu}_{de}}{a}~,
\end{equation}
where $\beta$ is the coupling constant whose magnitude determines the strength of the interaction rate. Here we consider the Hubble parameter $\mathcal{H}$ to be a global variable without any perturbation. When $\beta <0$, it is clear from equations (\ref{cons1}) and (\ref{cons2}) that DM redshifts faster than $a^{-3}$ while DE redshifts slowly. This is physically problematic as more of the DM is expected to be transferred to the DE budget in the late time, rather than in the beginning. For $\beta>0$, this problem is avoided. As shown by Feng and Zhang in \cite{feng}, for an HDE model, this form of interaction is favoured by geometrical data. We consider $\beta$ to be a free parameter. Using $u^{\mu}_{de} = \frac{1}{a} \delta^{\mu}_{0}$ and equation (\ref{Q-def}) in equation (\ref{interaction}) the interaction term $Q$ is obtained as
\begin{equation}
Q= \frac{\beta \mathcal{H} \rho_{de}}{a}~.
\end{equation}
Since dark energy dominates at the present epoch, we assume $w_{de} < -\frac{1}{3}$. As the motivation of the present work is to investigate the perturbation for a model without a big rip singularity, we restrict $w_{de} \geqslant -1$. For the non-interacting case $\left(\beta=0\right)$, it is clear from (\ref{int-eos}) that for $w_{de} \rightarrow -1$ at $z=0$, $C \rightarrow \sqrt{\Omega_{de0}}$, $\Omega_{de0}$ being the value of the dark energy density parameter at the present epoch. The present value of $\Omega_{de0}$ is taken from the Planck data\cite{planck2018} and is close to $0.6834$  which yields $C= 0.8267\simeq 0.83 $. In the presence of interaction, $C$ and $\beta$ have a correlation. The big rip singularity can be avoided if the interaction rate, $\beta$ lies between 
\begin{equation} \label{range}
-2 \sqrt{\frac{\Omega_{de0}}{C^2}} < \beta \leqslant 2- 2 \sqrt{\frac{\Omega_{de0}}{C^2}}~.
\end{equation}
The numerical values of $C$ and $\beta$ can be further constrained from other physical quantities like the deceleration parameter, $q$. For the IHDE model, $q$ depends on the parameters $C$ and $\beta$. In the subsequent part of this section, we will see the effect of interaction on the different physical quantities and try to constrain the parameter space for $C$ and $\beta$. 
\begin{figure}[!h]
  \centering
  \includegraphics[width=0.7\textwidth]{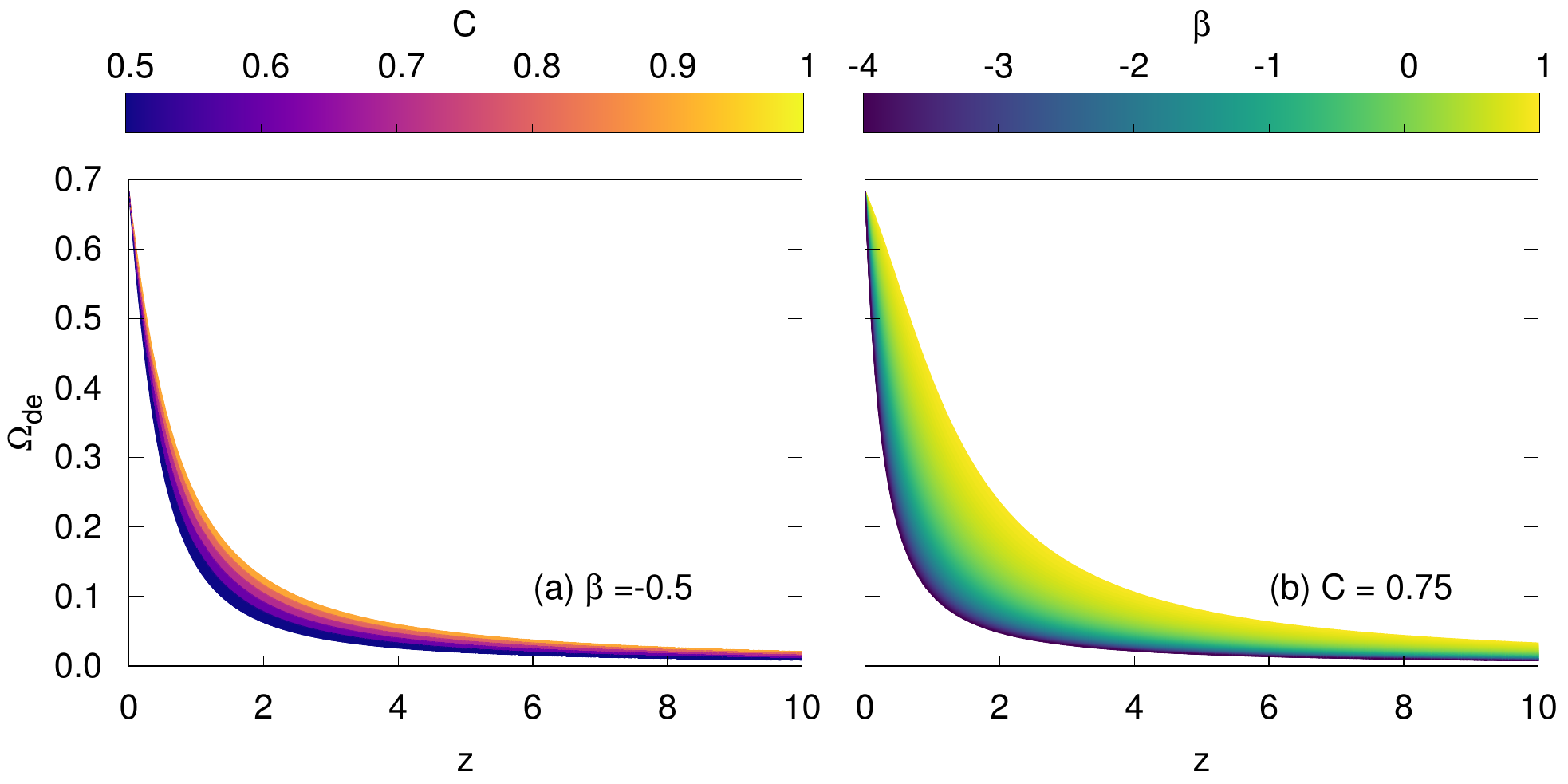}
\caption{(a) shows plot of $\Omega_{de}$ against $z$ for $C$ ranging from $0.5$ to $1.0$ and $\beta=-0.5$. (b) shows plot of $\Omega_{de}$ against $z$ for $\beta$ ranging from $-4.0$ to $1.0$ and $C=0.75$~.}\label{omegadez}
\includegraphics[width=0.7\textwidth]{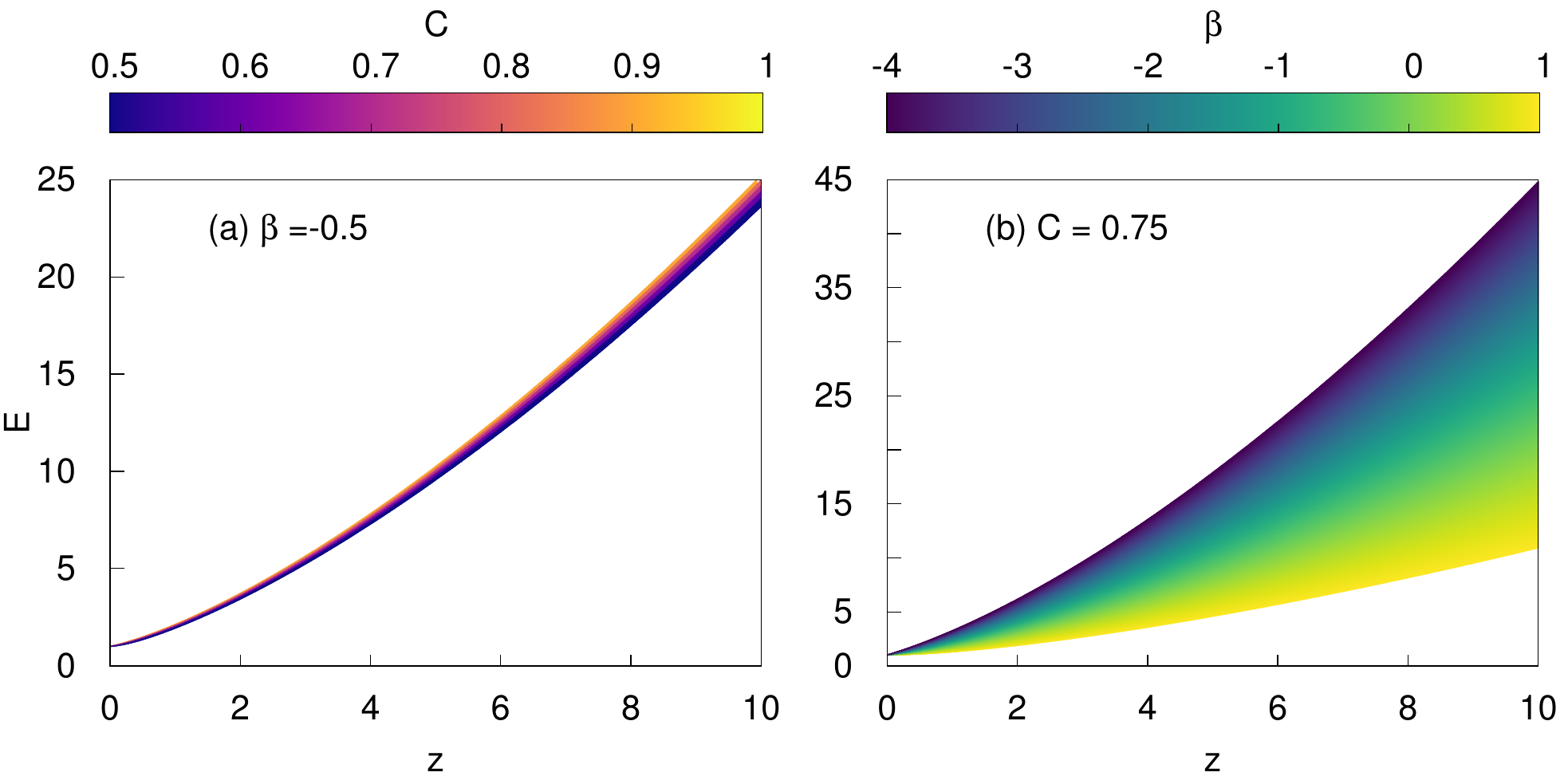}
\caption{(a) shows plot of $E$ against $z$ for $C$ ranging from $0.5$ to $1.0$ and $\beta=-0.5$. (b) shows plot of $E$ against $z$ for $\beta$ ranging from $-4.0$ to $1.0$ and $C=0.75$~.}\label{hz}
\end{figure}
Figures \ref{omegadez}, \ref{hz} and \ref{wz} show the evolution of the dark energy density parameter $\Omega_{de}$, the dimensionless Hubble parameter $E$ and the dark energy EoS parameter $w_{de}$ respectively with redshift $z$ for different values of $C$ and $\beta$. For figures \ref{omegadez} and \ref{hz}, equations (\ref{int-omega}) and (\ref{int-hubble}) are solved numerically from $z=0$ to $z=1100$ with the conditions that $\Omega_{de}(z=0) = \Omega_{de0}$ and $E(z=0) = 1$. For the effect of interaction, two cases are considered, (i) $C$ ranges from $0.5$ to $1.0$ (shades are darker as $C$ goes towards $0.5$) with $\beta$ taken as $-0.5$ and (ii) $\beta$ ranges from $-4$ to $1$ (shades are darker as $\beta$ to $-4$) with $C$ fixed at $0.75$. In figure \ref{omegadez}, $\Omega_{de}$ rises from nearly zero in the past to $0.68 \left(\sim 0.7\right)$ at present. For fixed non-zero interaction, the higher the value of $C$, the more gentle is the rise (figure \ref{omegadez}a). Similarly, for a fixed value of $C =0.75$, higher the value of $\beta$ the more gentle is the slope (figure \ref{omegadez}b). In figure \ref{hz}a, $E$ does not vary much in the past for different values of $C$ but in figure \ref{hz}b, $E$ varies quite remarkably for different values of $\beta$ --- the smaller the value of $\beta$ the higher the gradient.
\begin{figure}[!h]
  \centering
\includegraphics[width=0.7\textwidth]{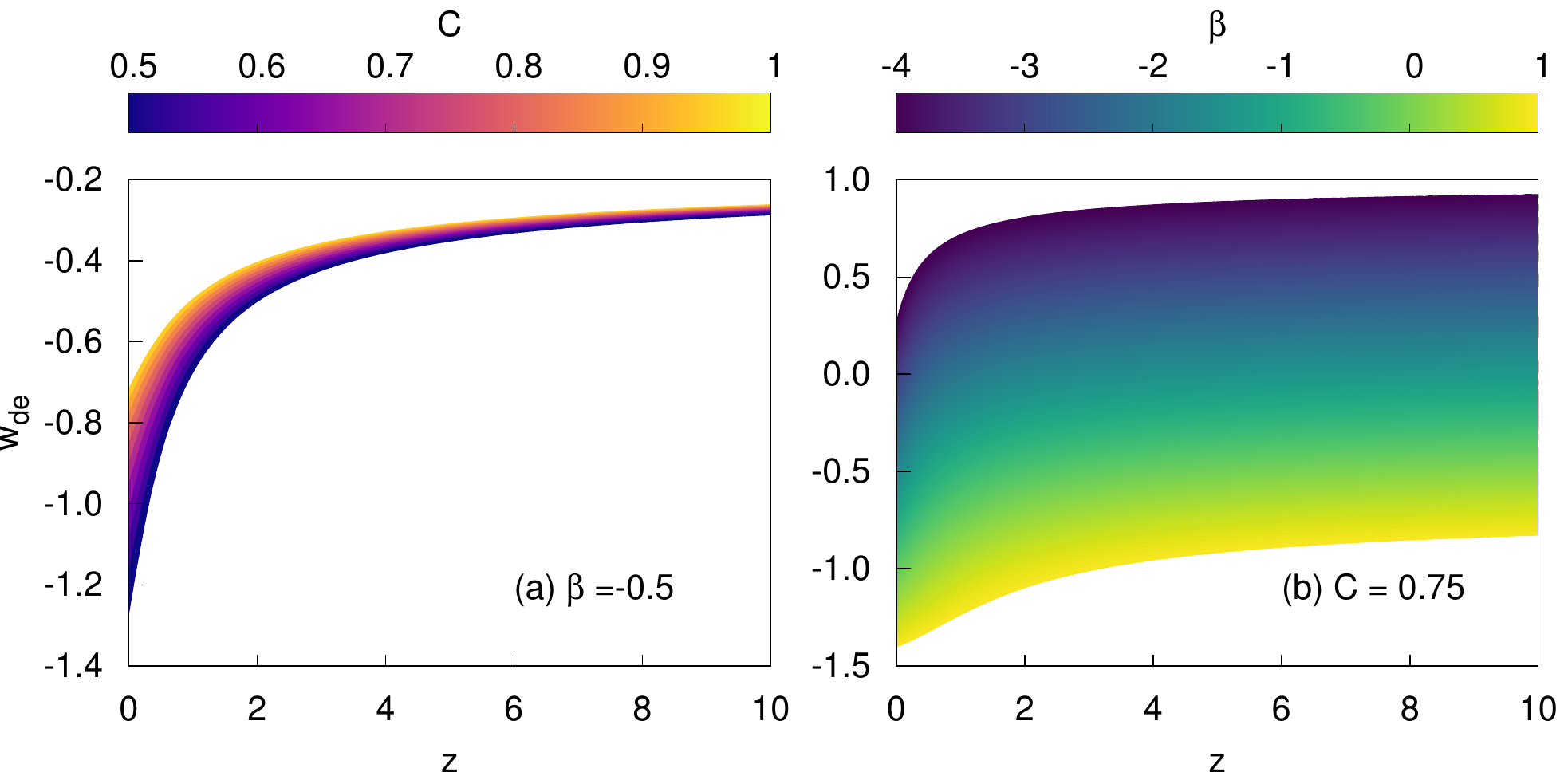}
\caption{(a) shows plot of $w_{de}$ against $z$ for $C$ ranging from $0.5$ to $1.0$ and $\beta=-0.5$. (b) shows plot of $w_{de}$ against $z$ for $\beta$ ranging from $-4.0$ to $1.0$ and $C=0.75$. For both of these plots $\Omega_{de}=0.68$ at $z=0$. Here $w_{de}$ ranges from nearly $-1.8$ to $0.4$ but only $-0.33<w_{de}\leqslant -1.0$ is considered for further calculation.} \label{wz}
\includegraphics[width=0.7\textwidth]{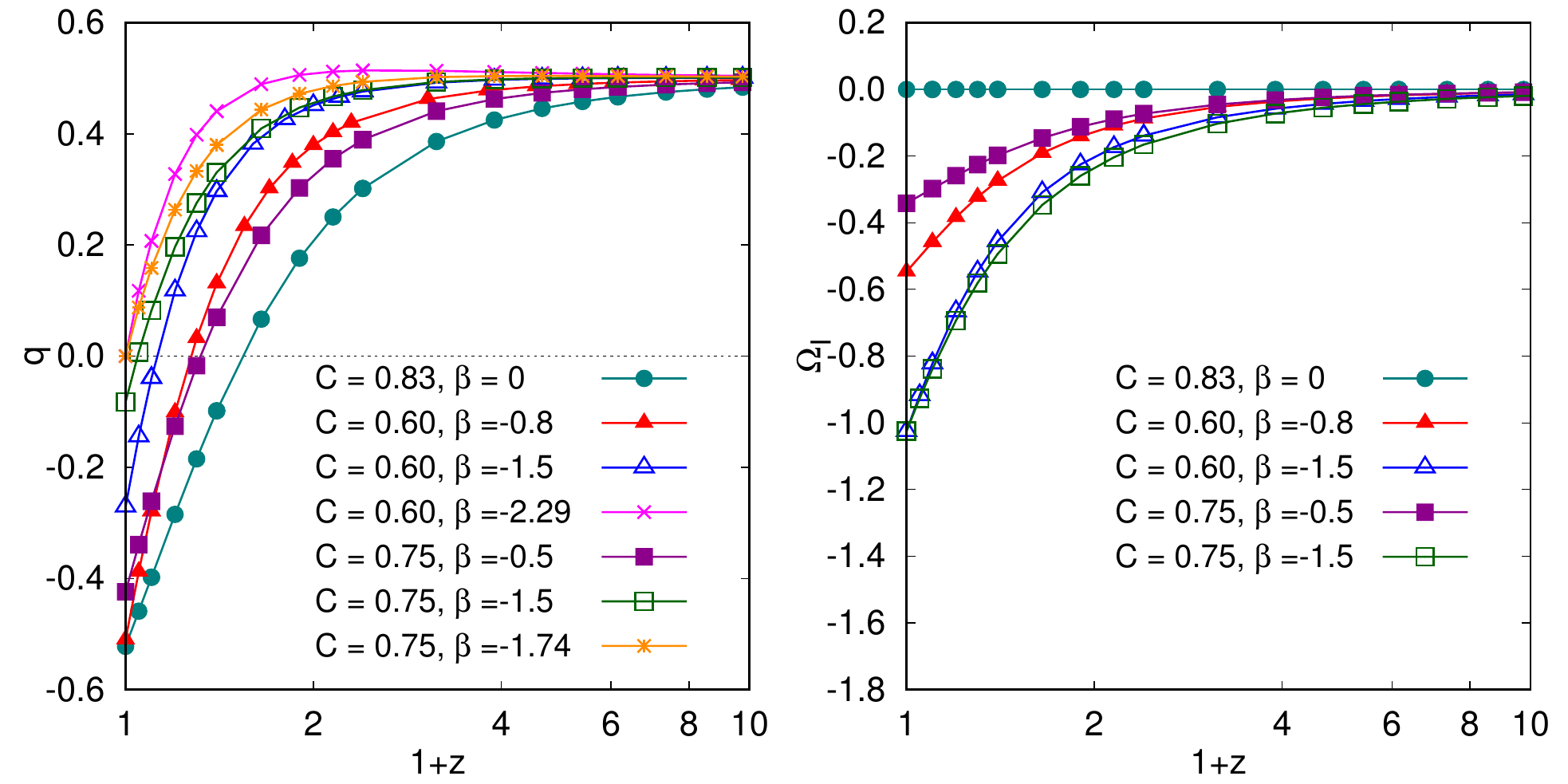}
\caption{(a) shows plot of $q$ against $\left(1+z\right)$ and (b) shows the variation of  $\Omega_I$ against $\left( 1+z\right)$ in logarithmic scale for different values of $C$ and $\beta$. The lines with solid circles corresponds to $C = 0.83$ and $\beta = 0$. The lines with triangles correspond to the interacting models with $C =  0.6$, $\beta = -0.8$ (solid) and $\beta = -1.5$ (hollow). The lines with squares correspond to the interacting models with $C =  0.75$, $\beta = -0.5$ (solid) and $\beta = -1.5$ (hollow). In figure (a) the lines with crosses ($C = 0.6$) and stars ($C = 0.75$) corresponds to the values of $\beta$ for which there is no acceleration at present.}\label{qz}
\end{figure}
 
For figures \ref{wz} a and \ref{wz} b, the solution of equation (\ref{int-omega}) is used in equation (\ref{int-eos}). For any value of $C$, $w_{de}$ increases and asymptotically reaches $-0.33$ at higher redshifts when $\Omega_{de} \sim 0$. In both the figures $w_{de}$ varies from nearly $-1.8$ to $0.4$ but only the parameter values for which $w_{de}$ lies between $-1$ and $-0.33$ are considered for further calculation. From figure \ref{wz}a, the allowed region for $C$ is constrained as $C \in\left[0.6,0.75\right]$. From figure \ref{wz}b and equation (\ref{range}) the allowed regions of $\beta$ are $\beta \in \left(-0.8 ,-1.5\right)$ (for $C=0.6$) and $\beta \in \left(-0.5 ,-1.5\right)$ (for $C=0.75$). 

Figures \ref{omegadez}, \ref{hz} and \ref{wz} are available in the literature, such as in \cite{mehrabi}, with the major difference that they are all without an interaction between the dark matter and dark energy unlike the present work. Anyway, we do require these figures to constrain $C$ and $\beta$ properly.

Figure \ref{qz} a shows the variation of the deceleration parameter $q$ with $\left(1+z\right)$ in logarithmic scale for different sets of $C$ and $\beta$. In all the cases, $q$ increases with $z$ and asymptotically approaches $0.5$ at higher redshifts. As seen from figure \ref{qz} a, for a fixed value of $C$ ($0.6$ and $0.75$), the smaller the value of $\beta$ ($-2.0$ and $-1.5$ respectively), the more recent is the transition from decelerated to accelerated Universe. For higher values of the coupling constant $\beta$ ($-0.8$ and $-0.5$), $q$ is nearly equal to $-0.5$ at present. Clearly, smaller values of $\beta (<-1.5)$ (larger magnitudes) will give decelerated Universe at present. Thus future event horizon as IR cut-off does not necessarily ensure accelerated Universe in the presence of an interaction. For non-interacting case, acceleration comes naturally as said by Li\cite{lim}. This figure brings out some new features, like it puts a limit on the strength of interaction. For achieving an accelerated model,  $\beta$ should be greater than $-1.741$ (for $C = 0.75$) or greater than $-2.292$ (for $C = 0.6$).

Figure \ref{qz}b shows the variation of the dimensionless interaction term $\Omega_{I}$ with with $\left(1+z\right)$ in logarithmic scale. For any pair of $C$ and $ \beta \left( \neq 0\right)$ the interaction term is nearly zero at higher redshifts and then starts to increase in magnitude with decrease in $z$. Since $\Omega_I = \beta \Omega_{de}$, the interaction increases in magnitude as the dark energy density parameter increases (figure \ref{omegadez}). With $\beta< 0$, from the definition of $\Omega_I$, we can see $Q<0$, which means energy is transferred from DM to DE. Thermodynamically energy transfer should be from DE to DM following Le Ch\^{a}telier--Braun principle as shown by Pav\'{o}n and Wang \cite{pavon2009}. In case of an HDE model negative $\beta$ (DM $\rightarrow$ DE) is slightly favoured by the data as shown by Zhang \etal\cite{zhang12}.

\section{The perturbation} \label{sec:pert}

In what follows, a scalar perturbation of the metric (\ref{metric}) is considered. In longitudinal gauge, the perturbed metric takes the form
\begin{equation} \label{metric2}
ds^2= a^2\left[-\left(1+2\Phi\right)d \eta^2 +\left(1-2\Psi \right)dx^i dx^j\right].
\end{equation}
Here $\Phi\left( \eta, \mathbf{x} \right)$ and $\Psi\left( \eta, \mathbf{x} \right)$ are the gauge-invariant variables, known as the Bardeen's potential \cite{bardeen}. In absence of any anisotropic stress, longitudinal gauge becomes identical to Newtonian gauge, as discussed in  \cite{lifshitz}, making $\Phi=\Psi$. Perturbations in the form of $\delta \rho_A\left( \eta, \mathbf{x} \right)$, $\delta p_A\left( \eta, \mathbf{x} \right)$, $\delta u_{\left(A\right)\mu} \left( \eta, \mathbf{x} \right)$ and $\delta Q_{\left(A\right)\mu}\left( \eta, \mathbf{x} \right)$ are added to the expression for energy-momentum tensor (\ref{stressA}). The components of the $4$-velocity perturbation of the fluid `A' are $\delta u_{\left(A\right)0} = -a \Phi$ obtained from the normalisation condition and $\delta u_{\left(A\right)i}= a \partial_i v_A$, $v_A$ being the peculiar velocity potential. Similarly, the total $4$-velocity perturbation $\delta u_\mu \left( \eta, \mathbf{x} \right)$ has the components $\delta u_\mu = -a \left(\Phi, \partial_i v \right)$ such that
\begin{equation}  \label{total-pert}
\left( \rho+ p\right) \theta = \sum_A \left(\rho_A +p_A\right)\theta_A~,
\end{equation}
with $\theta= -k^2 v $ being the divergence of the total fluid velocity \cite{ma} and $k$ is the wave number in the corresponding Fourier mode. For an individual fluid `A', the divergence of the corresponding velocity is $\theta_A= - k^2 v_A$. The perturbed energy-momentum transfer function $\delta Q_{\left(A\right) \mu}$ is split relative to the total $4$-velocity $u_\mu$ \cite{kodama, valiviita2008jcap} as
\begin{equation} \label{Q-pert-def}
\delta Q_{\left(A\right) \mu} = \delta Q_A u_\mu + F_{\left(A\right)\mu}, \hspace{1cm} u^\mu F_{\left(A\right)\mu} =0,
\end{equation}
where $\delta Q_A$ is the perturbation in the energy transfer rate, $F_{\left(A\right)\mu}= a \left( 0,\partial_i f_A\right)$ is the perturbation in the momentum density transfer rate and $f_A$ is the momentum transfer potential. It is clear from equation (\ref{condition}) that
\begin{equation}
\sum_A \delta Q_A =\sum_A f_A =0 \label{pert-condition}~.
\end{equation}

Writing the perturbation in Fourier modes, the $(00)$, $(0i) \equiv (i0)$ and $(ij)$ components of the Einstein field equations up to the first order in perturbation will read as\cite{mukhanov},
\begin{eqnarray}
&2\left[-3 \mathcal{H} \left( \mathcal{H} \Phi +\Phi^\prime \right) -k^2 \Phi \right]=\frac{3\mathcal{H}^2}{\rho}\delta \rho, &  \label{g002}\\
&2 k^2\left( \mathcal{H}\Phi +\Phi^\prime\right)=-\frac{3\mathcal{H}^2}{\rho}\left(p+\rho\right)\theta, & \label{g0j2}\\
& \hspace{-0.8cm}-2\left[ \left(2\mathcal{H}^\prime +\mathcal{H}^2\right) \Phi + 3\mathcal{H}\Phi^\prime +\Phi^{\prime \prime} \right] \delta^i{}_j=-\frac{3\mathcal{H}^2}{\rho}\delta p \delta^i{}_j. &\label{gij2}
\end{eqnarray}
The temporal and spatial parts of the first order in perturbation of the energy balance equation of the  fluid `A' \cite{valiviita2008jcap} are
\begin{eqnarray}
\begin{split}
\delta \rho^\prime_A + \left( \rho_A + p_A\right)\left(\theta_A -3 \Phi^\prime \right)& + 3 \mathcal{H} \left( \delta\rho_A +\delta p_A\right)\\
 =&aQ_A\Phi +a \delta Q_A , \label{e1}
\end{split}\\
\begin{split}
\left[\theta_A\left( \rho_A + p_A \right)\right]^\prime  &+ 4 \mathcal{H} \theta_A\left( \rho_A  + p_A \right)- k^2 \delta p_A\\
-k^2& \Phi\left( \rho_A  + p_A \right)=-k^2 a f_A+aQ_A \theta. \label{m1}
\end{split}
\end{eqnarray}
 
For an adiabatic perturbation in interacting fluids, the pressure perturbation $\delta p_A$ depends on $\delta \rho_A$ as well as on the interaction term $Q_A$ as\cite{waynehu, valiviita2008jcap, bean} 
\begin{equation}\label{pert-p}
\delta p_A=c_{s,A}^2 \delta \rho_A+\left(c_{s,A}^2-c_{a,A}^2\right)\left[3 \mathcal{H}\left(1+w_A\right)\rho_A -a Q_A\right]\frac{\theta_A}{k^2},
\end{equation}
where 
\begin{equation}\label{ca2}
c_{a,A}^2=\frac{p_A^\prime}{\rho_A^\prime}=w_A+\frac{w_A^\prime}{\rho_A^\prime/p_A^\prime}
\end{equation}
is the adiabatic speed of sound of `A' fluid and $c_{s,A}^2$ is the effective speed of sound of `A' fluid, defined as
\begin{equation}\label{cs2}
c_{s,A}^2=\frac{\delta p_{A}}{\delta \rho_{A}} \bigg \rvert_{rest,A},
\end{equation}
is the ratio of pressure fluctuation to density fluctuation in the rest frame of  fluid `A'.

As shown in \cite{mehrabi1478}, $c_{s,de}^2$ plays a significant role in DE clustering and hence DM clustering. When $c_{s,de}^2 \simeq 1$, the pressure perturbation should suppress any growth in DE perturbation whereas when $c_{s,de}^2 \ll 1$, DE perturbation should grow like that of DM.  It is shown in \cite{mehrabi1478, batista2013, batista2017} that DE can cluster like DM when $c_{s,de}^2=0$. Here for the case of CDM and IHDE we consider that $0 \leqslant c_{s,de}^2 \leqslant 1$ and study the evolution of DM and DE perturbations in detail for $c_{s,de}^2 =0$. Then we compare some of the results for different non-zero values of $c_{s,de}^2$. It deserves mention that the effective sound speed $c_{s,de}^2$, defined in equation (\ref{cs2}) is different from the adiabatic sound speed $c_{a,de}$, defined in equation (\ref{ca2}).
\section{Evolution of the density contrasts} \label{sec:evolution}

We shall now frame the differential equations for density contrasts for both DM and DE that determine their evolution with redshift. For that, we need to know the perturbation in the interaction term itself. From equations (\ref{Q-pert-def}) and (\ref{pert-condition}), it follows that
\begin{eqnarray}
\delta Q_{m}= -\delta Q_{de}  = \frac{\beta \mathcal{H} \delta \rho_{de} }{a}, \label{pert-int1}  \\
 f_{m} = - f_{de} = \frac{\beta \mathcal{H} \rho_{de}\left(\theta -\theta_{de}\right)}{a k^2}. \label{pert-int2}
\end{eqnarray}

The density contrasts of CDM and IHDE are $\delta_m =\frac{\delta \rho_m}{\rho_m}$ and $\delta_{de} =\frac{\delta \rho_{de}}{\rho_{de}}$ respectively. Using (\ref{pert-int1}) and (\ref{pert-int2}), the equations (\ref{e1}) and (\ref{m1}) for CDM and IHDE can be written respectively as 
\begin{eqnarray}
\delta_{m}^\prime+\theta_m-3 \Phi^\prime &=& \beta \mathcal{H}\frac{\rho_{de}}{\rho_m}\left(\Phi-\delta_m +\delta_{de}\right), \label{e2m}\\
\theta_m^\prime+\mathcal{H}\theta_m-k^2 \Phi&=& -\beta \mathcal{H}\frac{\rho_{de}}{\rho_m}\left(\theta_m -\theta_{de}\right), \label{m2m}
\end{eqnarray}
\begin{equation}
\begin{split}
\delta_{de}^\prime +3 \mathcal{H}\left(c_{s,de}^{2}-w_{de}\right)\delta_{de}+\left(1+w_{de}\right)\left(\theta_{de}-3\Phi^\prime\right)+3\mathcal{H}\left[3 \mathcal{H} \left(1+w_{de}\right)\left(c_{s,de}^{2}-w_{de}\right)\right]&\frac{\theta_{de}}{k^2} +3\mathcal{H} w_{de}^\prime\frac{\theta_{de}}{k^2} \\
=-\beta \mathcal{H}&\left[\Phi+3 \mathcal{H}\left(c_{s,de}^{2}-w_{de}\right)\frac{\theta_{de}}{k^2}\right], \label{e2de}
\end{split}
\end{equation}
\begin{equation}
\theta_{de}^\prime+\mathcal{H}\left(1-3c_{s,de}^{2}\right)\theta_{de}-k^2\Phi-\frac{k^2\delta_{de}c_{s,de}^{2} }{\left(1+w_{de}\right)}=\frac{\beta \mathcal{H}}{\left(1+w_{de}\right)}\left(-c_{s,de}^{2}\theta_{de}\right). \label{m2de}
\end{equation}

Eliminating $\theta_m$ from equations (\ref{e2m}), (\ref{m2m}) and $\theta_{de}$ from equations (\ref{e2de}), (\ref{m2de}),  the coupled differential equations for CDM and IHDE are obtained respectively in terms of redshift as
\begin{eqnarray}
\mathbf{C^{(m)}_1} \frac{\partial^2 \delta_m}{\partial z^2}+ \mathbf{C^{(m)}_2} \frac{\partial\delta_m}{\partial z} +\mathbf{C^{(m)}_3} \delta_m + \mathbf{C^{(m)}_4} \frac{\partial^2 \delta_{de}}{\partial z^2}+\mathbf{C^{(m)}_5} \frac{\partial \delta_{de}}{\partial z}+ \mathbf{C^{(m)}_6} \delta_{de}+ \mathbf{C^{(m)}_7} \frac{\partial^2 \Phi}{\partial z^2}+ \mathbf{C^{(m)}_8} \frac{\partial \Phi}{\partial z} + \mathbf{C^{(m)}_9} \Phi \hspace{-0.3cm} &=&\hspace{-0.3cm}0~, \label{finaldm}\\
\mathbf{C^{(de)}_1} \frac{\partial^2 \delta_{de}}{\partial z^2}+ \mathbf{C^{(de)}_2} \frac{\partial\delta_{de}}{\partial z} +\mathbf{C^{(de)}_3} \delta_{de} + \mathbf{C^{(de)}_4} \frac{\partial^2 \delta_{m}}{\partial z^2}+\mathbf{C^{(de)}_5} \frac{\partial \delta_{m}}{\partial z} + \mathbf{C^{(de)}_6}\delta_m+ \mathbf{C^{(de)}_7} \frac{\partial^2 \Phi}{\partial z^2}+ \mathbf{C^{(de)}_8} \frac{\partial \Phi}{\partial z} + \mathbf{C^{(de)}_9} \Phi \hspace{-0.3cm} &=&\hspace{-0.3cm}0~. \label{finalde}
\end{eqnarray}

The coefficients $\mathbf{C_1}$,  $\mathbf{C_2}$, $\mathbf{C_3}$, $\mathbf{C_4}$, $\mathbf{C_5}$, $\mathbf{C_6}$, $\mathbf{C_7}$, $\mathbf{C_8}$ and $\mathbf{C_9}$ are given in the Appendix \ref{appendix}. The coefficients $\mathbf{C^{(de)}_4}$, $\mathbf{C^{(de)}_5}$ and $\mathbf{C^{(de)}_6}$ are zero in equation (\ref{finalde}) indicating that the evolution of DE is not directly affected by DM fluctuation but the converse is not true. The coefficients $\mathbf{C_7}$, $\mathbf{C_8}$ and $\mathbf{C_9}$ are non zero in both the equations (\ref{finaldm}) and (\ref{finalde}) which implies that the potential $\Phi$ will affect the evolution of both DM and DE density contrasts. The evolution of $\Phi$ is governed by the equation (\ref{g002}) and is not approximated by the Poisson equation. The equations (\ref{finaldm}) and (\ref{finalde}) along with the equation (\ref{g002}) are solved numerically to find the evolution of the density contrasts of the CDM and IHDE. In order to do that, in the matter-dominated era, i.e. from $z_{in} = 1100$, it is assumed that $\Phi\left(z_{in}\right)=\mbox{constant}=\phi_0$ and $\Phi^\prime\left(z_{in}\right)=0$. It is also assumed that $\Omega_{m}\left(z_{in}\right) >> \Omega_{de}\left(z_{in}\right)$ so that the term with the ratio $\frac{\Omega_{de}\left(z_{in}\right)}{\Omega_m\left(z_{in}\right)}$ can be neglected for $\delta_m\left(z_{in}\right)$ only. As discussed in \cite{abramo}, the initial conditions for $\delta_m$, $\delta^\prime_m$, $\delta_{de}$ and $\delta^\prime_{de}$ are
\begin{eqnarray}
\delta_{mi}&=&\delta_m\left(z_{in}\right) ~=~ -2 \phi_0\left[1+\frac{\left(1+z_{in}\right)^2 k_{in}^2}{3 H_{in}^2}\right]~,\\
\delta_{mi}^\prime &=&\frac{d \delta_m}{d z}\bigg \rvert_{\substack{z=z_{in}}}=-4 \phi_0 \frac{\left(1+z_{in}\right) k_{in}^2}{3 H_{in}^2} \left[1-\frac{3}{2}\left (1+\frac{w_{dei} \Omega_{dei}}{\Omega_{mi}}\right )\right ]~,\\
\delta_{dei} &=&\delta_{de}\left(z_{in}\right) ~=~ \frac{\delta_{mi}}{3-\beta \frac{\Omega_{dei}}{\Omega_{mi}}}\lbrace 3\left(1+w_{dei}\right)+\beta \rbrace ~,\\
\delta_{dei}^\prime &=&\frac{d \delta_{de}}{d z}\bigg \rvert_{\substack{z=z_{in}}}=\frac{3 \left(\frac{d w_{de}}{d z}\right)\Big \rvert_{\substack{z=z_{in}}}}{3-\beta \frac{\Omega_{dei}}{\Omega_{mi}}}+\frac{\delta_{mi}^\prime}{3-\beta \frac{\Omega_{dei}}{\Omega_{mi}}}\lbrace 3\left(1+w_{dei}\right)+\beta\rbrace +\delta_{mi}\lbrace 3\left(1+w_{dei}\right)+\beta\rbrace \frac{\beta \left[ \frac{d}{d z}\left(\frac{\Omega_{de}}{\Omega_m}\right)\right]_{\substack{z=z_{in}}}}{\left(3-\beta \frac{\Omega_{dei}}{\Omega_{mi}}\right)^2}~,\hspace{0.2cm}
\end{eqnarray}
where $H_{in}=H\left(z_{in}\right)$, $w_{dei}=w_{de}\left(z_{in}\right)$, $\Omega_{mi}=\Omega_m\left(z_{in}\right)$, $\Omega_{dei}=\Omega_{de}\left(z_{in}\right)$ and $k_{in}$ is the mode entering the Hubble horizon at $z_{in}$. The value of $k_{in}$ is taken as $\left(1+z\right)^{-1}H_{in}$. The numerical values for $H_{in}$, $w_{dei}$, $\Omega_{mi}$ and $\Omega_{dei}$ are obtained from the solutions of the equations (\ref{int-omega}), (\ref{int-hubble}) and (\ref{int-eos}). The value of $\phi_0$ is given by hand.

For the Fourier mode, $k$ in equations (\ref{finaldm}), (\ref{finalde}) and (\ref{g002}), the value is considered in the linear regime given by the galaxy power spectrum\cite{percival07}
\begin{equation} \label{k-linear}
0.01 h \hskip1ex \mbox{Mpc}^{-1} \lesssim k \lesssim 0.2 h \hskip1ex \mbox{Mpc}^{-1},
\end{equation}
where $h = \frac{H_0}{100 \hskip1ex \footnotesize{\mbox{km s}^{-1} \mbox{Mpc}^{-1}}}$ is the dimensionless Hubble parameter at the present epoch. The value of $H_0 = 67.27$ is taken from the Planck data\cite{planck2018}. For $k > 0.2 h \hskip1ex \mbox{Mpc}^{-1}$ (smaller scales), non-linear effects become prominent whereas for $k < 0.01 h \hskip1ex \mbox{Mpc}^{-1}$ observations are not very accurate. So we consider $k=0.1 h \hskip1ex \mbox{Mpc}^{-1}$ following \cite{mehrabi1478}. For our calculation we have considered $\phi_0 =10^{-5}$ and $c_{s,de}^2=0$. 

\begin{figure}[!h]
  \centering
\includegraphics[width=0.7\textwidth]{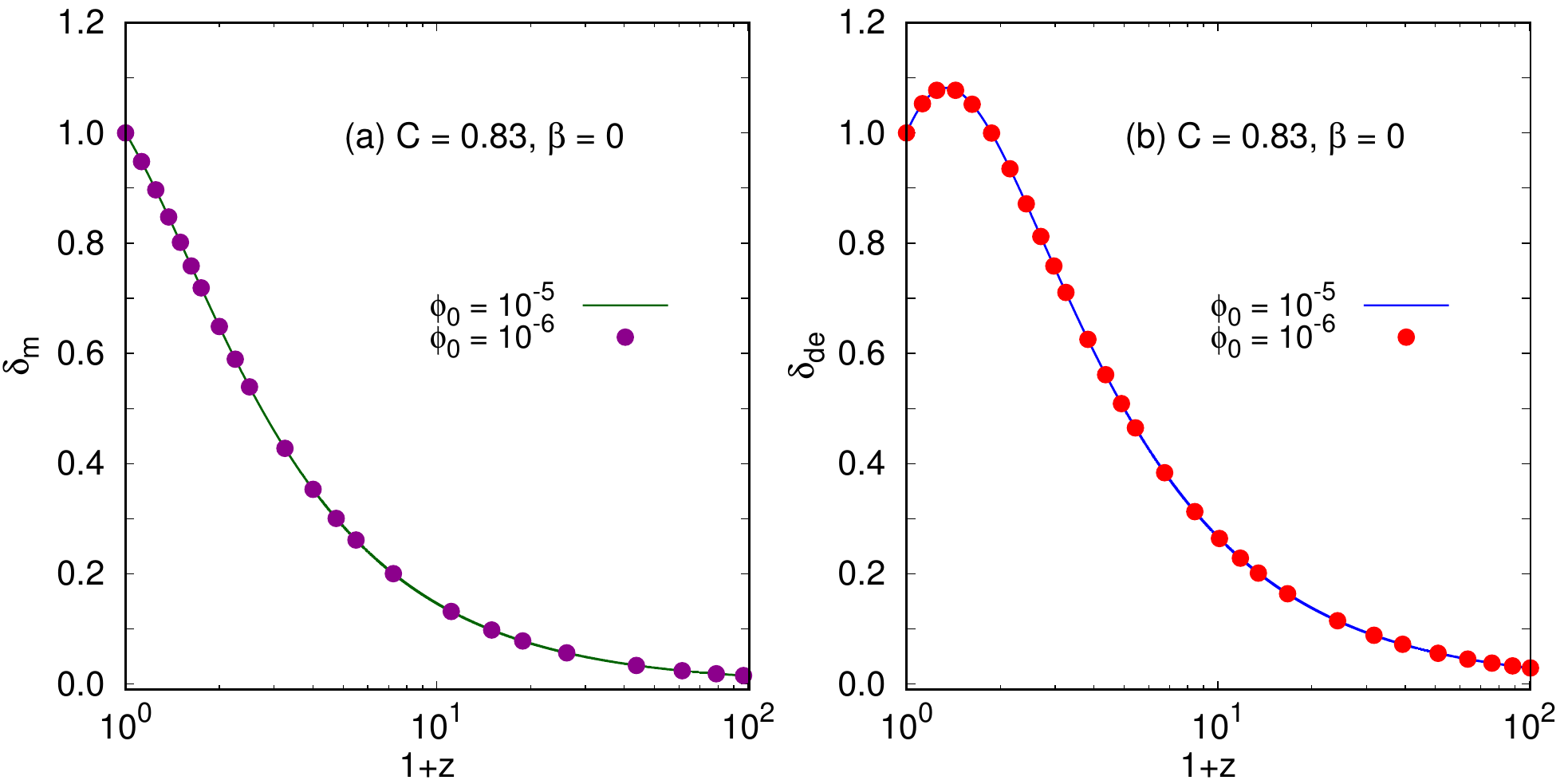}
\caption{(a) shows plot of $\delta_m$ against $\left(1+z\right)$ and (b) shows the plot of $\delta_{de}$ against $\left(1+z\right)$ in logarithmic scale for $C =0.83$ and $\beta = 0$. The line shows the variation of $\delta_m$ and $\delta_{de}$ for the initial condition, $\phi_0 = 10^{-5}$ and the solid circles represent the same corresponding to the initial condition, $\phi_0 = 10^{-6}$~.}\label{figmul0183}
  \centering
\includegraphics[width=0.7\textwidth]{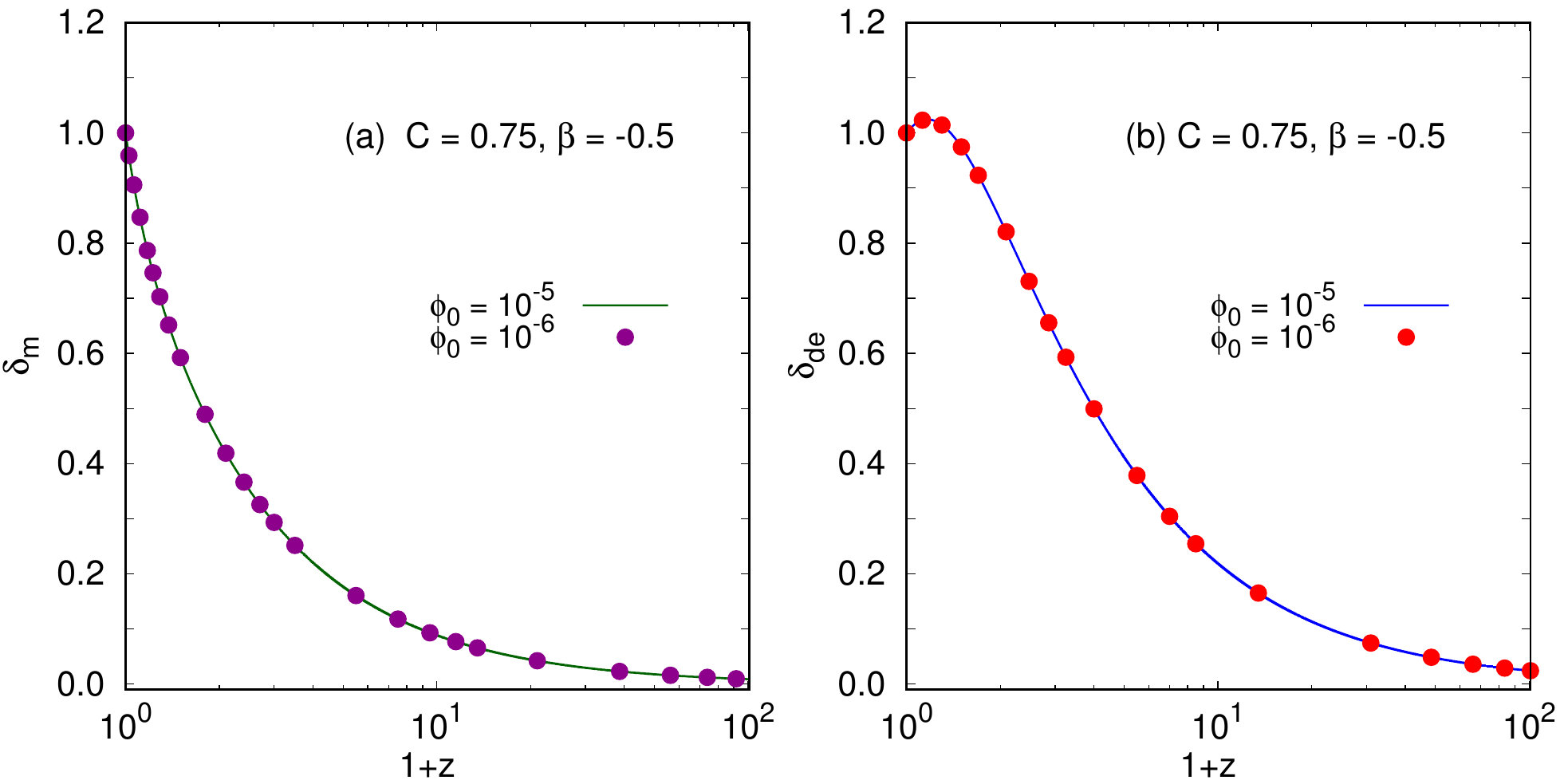}
\caption{(a) shows plot of $\delta_m$ against $\left(1+z\right)$ and (b) shows the plot of $\delta_{de}$ against $\left(1+z\right)$ in logarithmic scale for $C =0.75$ and $\beta =-0.5$. The line shows the variation of $\delta_m$ and $\delta_{de}$ for the initial condition, $\phi_0 = 10^{-5}$ and the solid circles represent the same corresponding to the initial condition, $\phi_0 = 10^{-6}$~.}\label{figmul0175}
\end{figure}

\begin{figure}[!h]
  \centering
\includegraphics[width=0.7\textwidth]{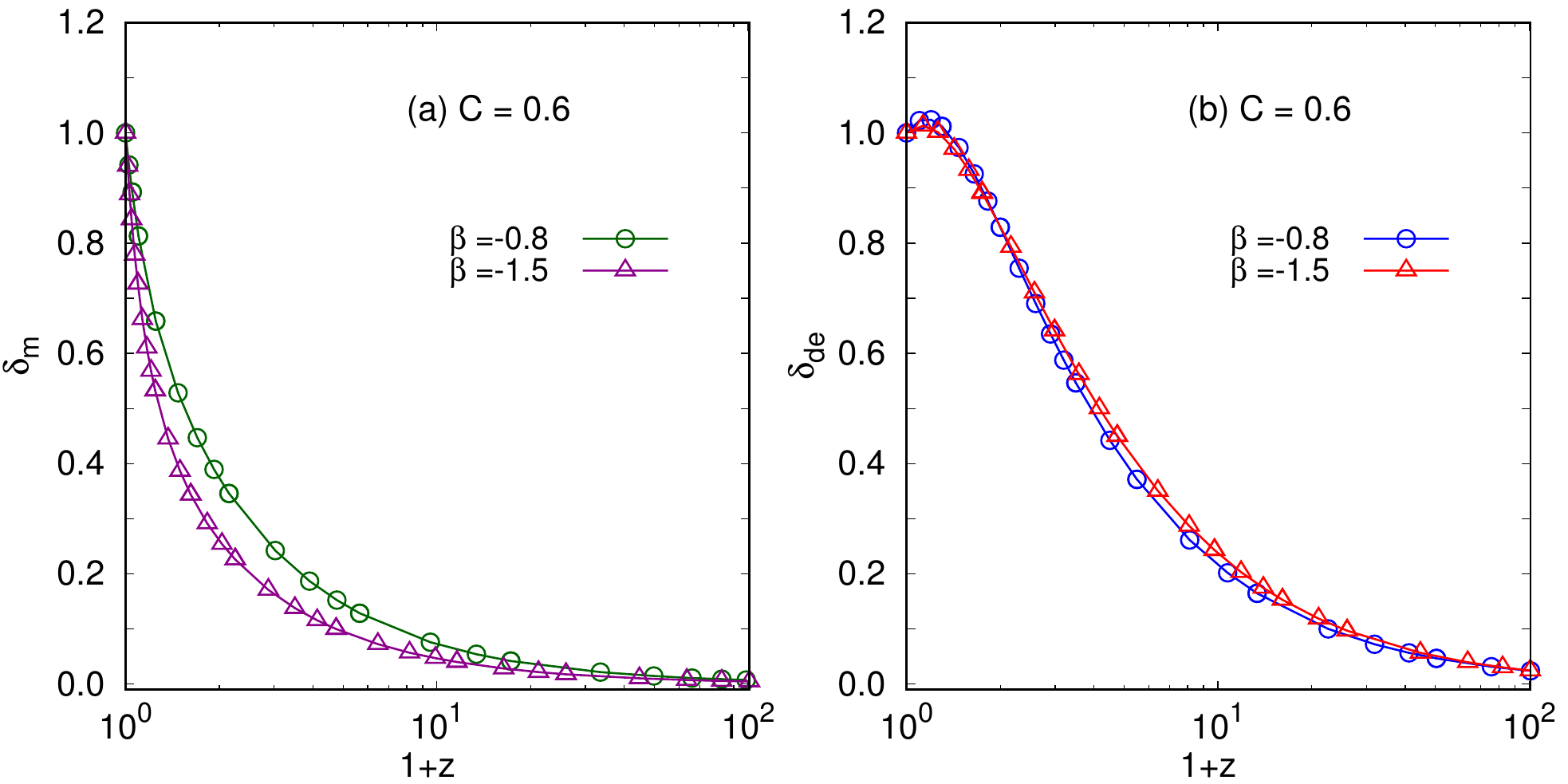}
\caption{(a) shows plot of $\delta_m$ against $\left(1+z\right)$ and (b) shows the plot of $\delta_{de}$ against $\left(1+z\right)$ in logarithmic scale for $C =0.6$ and two different values of $\beta$. The line with circles is for $\beta=-0.8$ and the line with triangles is for $\beta = -1.5$~.}\label{fig016all}
  \centering
\includegraphics[width=0.7\textwidth]{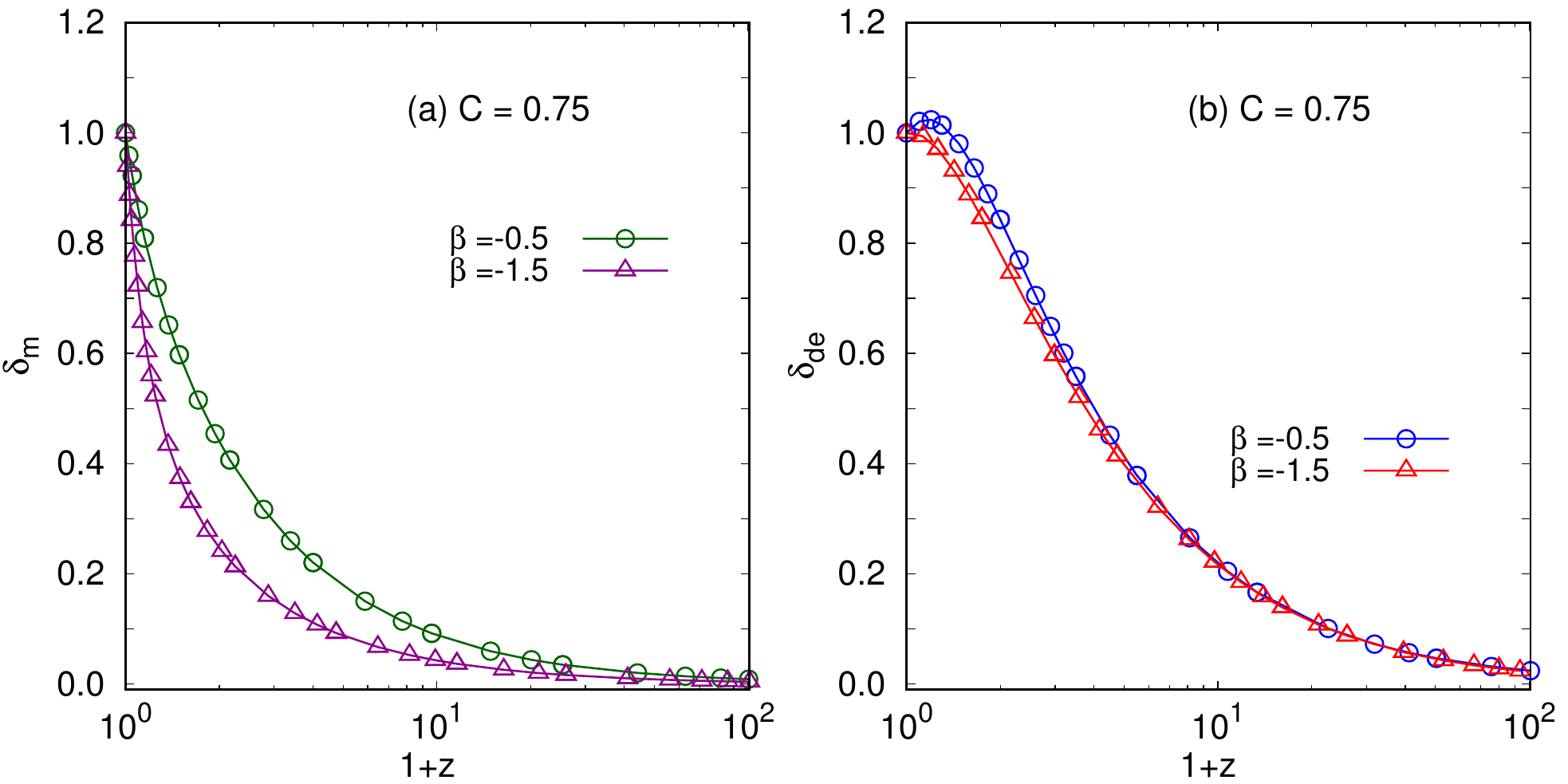}
\caption{(a) shows plot of $\delta_m$ against $\left(1+z\right)$ and (b) shows the plot of $\delta_{de}$ against $\left(1+z\right)$ in logarithmic scale for $C =0.75$ and two different values of $\beta$. The line with circles is for $\beta=-0.5$ and the line with triangles is for $\beta = -1.5$~.}\label{fig0175all}
\end{figure}

\begin{figure}[!h]
  \centering
    \includegraphics[width=0.7\textwidth]{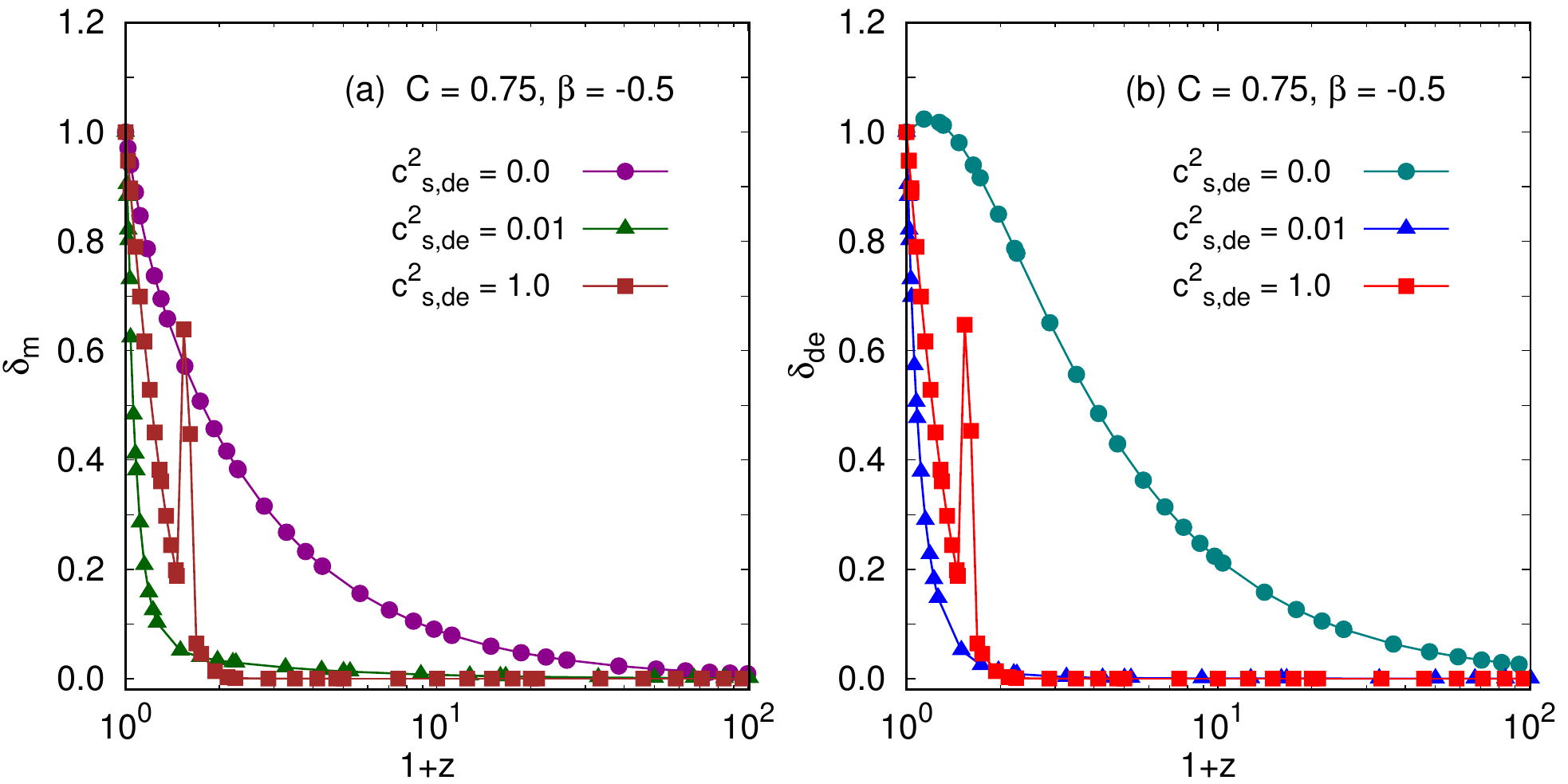}
\caption{(a) shows plot of $\delta_m$ against $\left(1+z\right)$ and (b) shows the plot of $\delta_{de}$ against $\left(1+z\right)$ in logarithmic scale for different values of $c_{s,de}^2$ with $C =0.75$ and $\beta=-0.5 $. The line with solid circles corresponds to $c_{s,de}^2 =0$, the line with solid triangles corresponds to $c_{s,de}^2 =0.01$ and the line with solid squares corresponds to $c_{s,de}^2 =1.0 $~.} \label{figcs}
  \centering
    \includegraphics[width=0.7\textwidth]{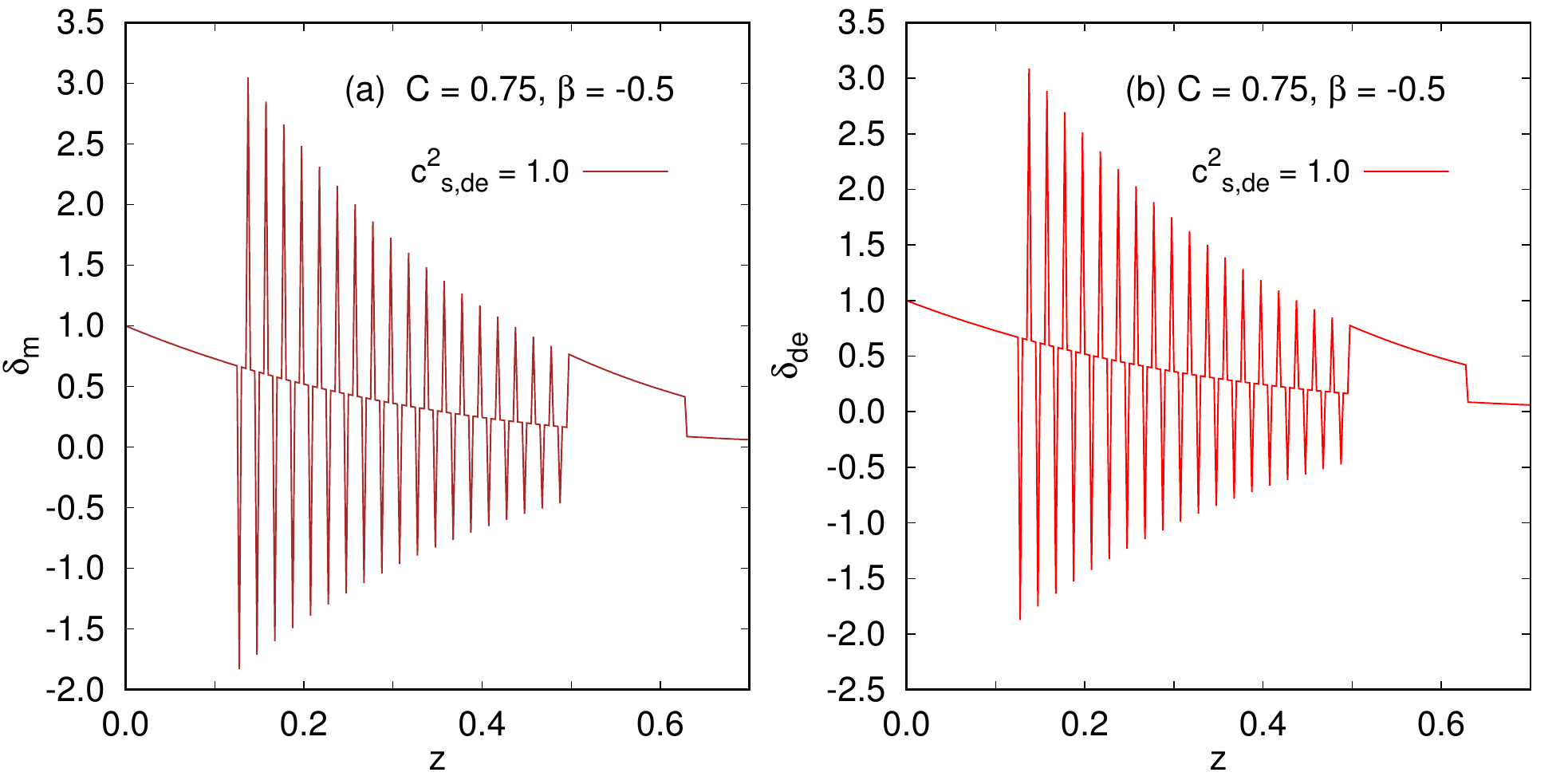}
\caption{(a) shows plot of $\delta_m$ against $z$ and (b) shows the plot of $\delta_{de}$ against $z$ from $z=0$ to $z=0.7$ for $c_{s,de}^2=1.0$ with $C =0.75$ and $\beta=-0.5$~.} \label{figcs1}
 \includegraphics[width=0.7\textwidth]{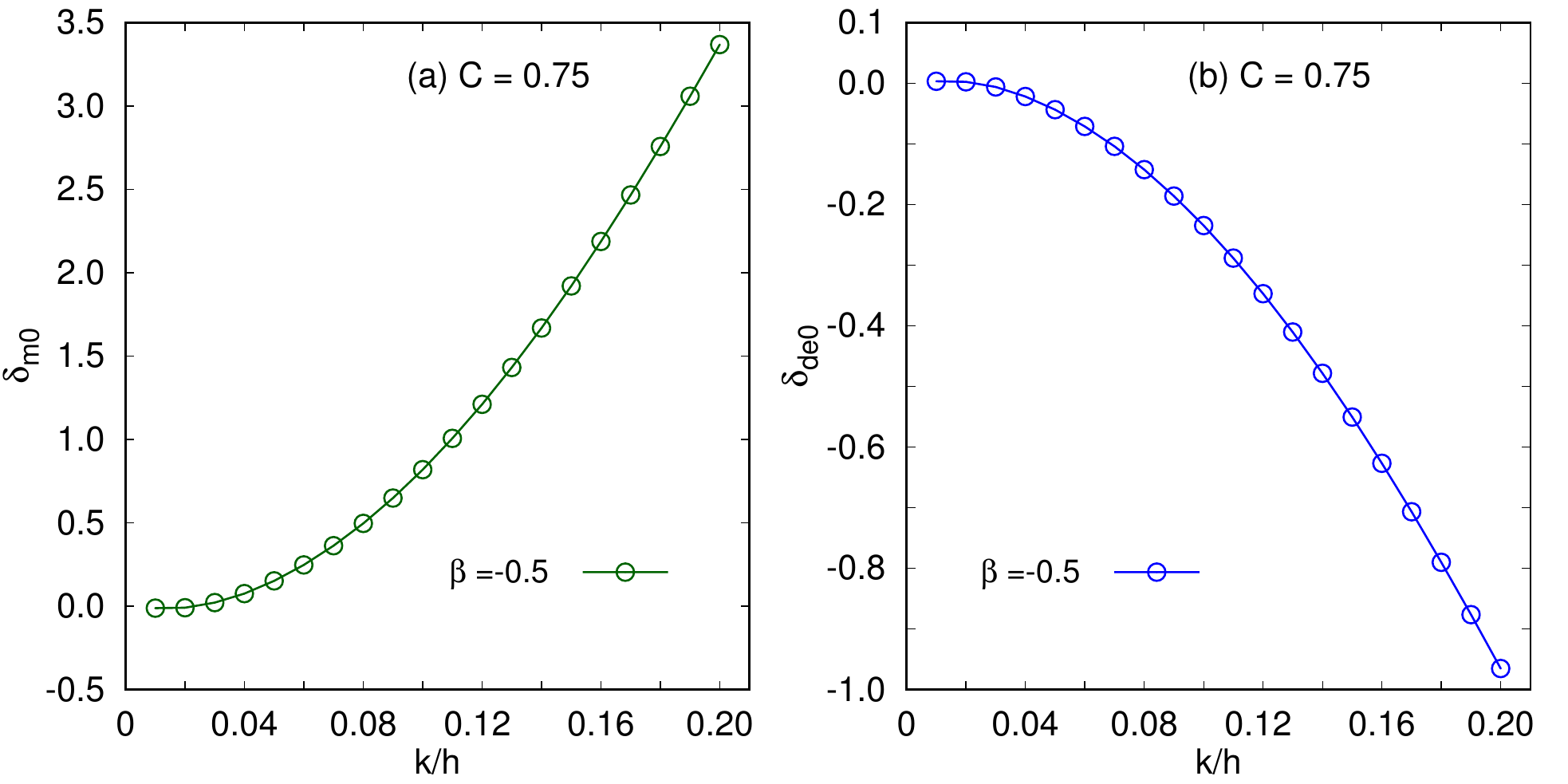}
\caption{(a) shows plot of $\delta_{m0}$ against $k$ and (b) shows the plot of $\delta_{de0}$ against $k$ at $z=0$ for $C =0.75$ and $\beta=-0.5$~.} \label{fig01k}
\end{figure}

The density contrast of DM, $\delta_m$ has over density (positive solution) while that of DE, $\delta_{de}$ has under density (negative solution). All the figures \ref{figmul0183}-\ref{figcs} are shown in logarithmic scale from $z=0$ to $z=100$ with $C=0.83$ and $\beta=0$ for the non-interacting case. For the interacting case, we have chosen $C=0.75$ and $\beta = -0.5$. In all the figures, the density contrast is scaled by its present value. To study the effect of interaction in the growth of the density contrasts we have considered different sets $C$ and $\beta$.

Figure \ref{figmul0183} shows the variation of $\delta_m$ and $\delta_{de}$ with $\left(1+z\right)$ in logarithmic scale for the non-interacting case for $\phi_0 = 10^{-5}$ and $\phi_0 = 10^{-6}$. When scaled by their respective present value, the nature of the growth of $\delta_m$ and $\delta_{de}$ is hardly sensitive to the value of $\phi_0$. This is clear from figures \ref{figmul0183} a and \ref{figmul0183} b. Figure \ref{figmul0175} shows the same for the interacting case with $C=0.75$ and $\beta =-0.5$. One can clearly see from figures \ref{figmul0183} a and \ref{figmul0175} a that the interaction ($\beta \ne 0$) makes the slopes different.   For the variation of $\delta_{de}$, we see that it first grows up to a maximum and then decreases to unity at $z=0$. The position as well as the height of this peak is different in the figures \ref{figmul0183} b and \ref{figmul0175} b. The presence of an interaction has decreased the height of the maximum and made the growth a little flat.

Figures \ref{fig016all} a and \ref{fig0175all} a show the variation of $\delta_m$ with $\left(1+z\right)$ in logarithmic scale for the same value of $C$ but different values of $\beta$. For $C=0.6$, as $\beta$ decreases from $-0.8$ to $-2.0$ the curve tends to become steeper and the growth becomes faster. This behaviour is similar in both the cases of $C=0.6$ and $C=0.75$.  For $\delta_{de}$ (figures \ref{fig016all} b and \ref{fig0175all} b), the change in slope for smaller $\beta$ is more prominent in smaller $C$ value. In figure \ref{fig016all} b, for $C=0.6$, $\delta_{de}$ for $\beta = -1.5$, changes faster than that for $\beta=-0.8$. Similarly in figure \ref{fig0175all} b, for $C=0.75$, $\delta_{de}$ for $\beta=-1.5$ changes faster than that for $\beta=-0.5$. The change in the direction of the growth rate takes place after the Universe starts accelerating and has a correlation with the deceleration parameter $q$ changing its sign. For figure \ref{fig016all} b,  the maximum of ${\delta}_{de}$ for $\beta = -1.5$ is at a slightly lower redshift than that for $\beta =-0.8$, and for figure \ref{fig0175all} b, the maximum of ${\delta}_{de}$ for $\beta =-1.5$ is at a lower redshift than that for $\beta =-0.5$. If $\beta$ is decreased below $-1.5$, no such correlation is seen.

Figure \ref{figcs} shows the variation of the density contrasts $\delta_m$ and $\delta_{de}$ for the different values of the effective speed of sound of dark energy perturbation, $c_{s,de}^2$. From the expression of $\delta p_{de}$ equation (\ref{pert-p}), we can see that the pressure perturbation not only depends on the product $c_{s,de}^2 \delta \rho_{de}$ but also on the background quantities like $w_{de}$, $ \rho_{de}$, $Q$ as well as the velocity perturbation through $\theta_{de}$. With $c_{s,de}^2=0$, the effect of the velocity perturbation present in the second term is prominent. The presence of the interaction $Q$ actually decreases this effect (figures \ref{figmul0183} b and \ref{figmul0175} b). In presence of the $c_{s,de}^2$ (i.e.\ any non-zero value), the effect of $\delta \rho_{de}$ comes into play. The growth is more steep if the proportion of $\delta \rho_{de}$ increases. When zoomed into smaller redshift region ($z=0.1275$ to $z=0.4875$), rapid oscillations are observed (figure \ref{figcs1}). 

Figure \ref{fig01k} shows the variation of $\delta_m$ and $\delta_{de}$ with $k/h$ at $z=0$.  In figure \ref{fig01k} a, for a given value of $C$ and $\beta$, as $k/h$ increases $\delta_{m0}$ increases --- the growth rate of DM over densities increase for smaller scales entering the horizon. Though the increase is not linear for $k/h$ less than $ \sim 0.1h$, but when scaled by $\delta_{m0}$, the growth rates of $\delta_m$ for different $k$-modes are independent of the $k$-modes. For figure \ref{fig01k} b, negative values of $\delta_{de}$ increases with larger scales. The change in slope in this case is also not linear for modes smaller than $\sim 0.1h$ and the growth of $\delta_{de}/\delta_{de0}$ for different modes are identical.

\section{Summary and Discussion} \label{sec:summary}

The primary motivation of the present work is to study the effect of interaction on density perturbation in the dark sector of the Universe in a Holographic dark energy model. Among various possibilities, we have chosen the future event horizon as the IR cut-off for the HDE model for which the Universe can accelerate even in the absence of an interaction. For the interaction between the DM and DE, we have chosen the interaction term to be proportional to the dark energy density $\rho_{de}$. The interaction term is of the form $Q=\frac{\beta \mathcal{H}\rho_{de}}{a}$, in which the dependence on cosmic time comes through the global expansion rate, the Hubble parameter $\mathcal{H}$. The coupling constant $\beta$ determines the strength of the interaction as well as the direction of the energy flow. No interaction between DE and DM is characterised by $\beta=0$. We restricted the model parameters $C$ and $\beta$ in such a way that at present the DE EoS parameter, $w_{de}$ is sufficiently negative to produce the late time acceleration but can avoid the ``phantom menace'' ($w_{de}< -1$).

The set of coupled second order differential equations for the density contrasts for the CDM ($\delta_m$) as well as the HDE ($\delta_{de}$) were obtained in the Newtonian gauge. Adiabatic initial conditions were used with the assumption that the DE density parameter is small compared to the DM density parameter ($\frac{\Omega_{de}}{\Omega_m} \ll 1$) at the onset of the matter dominated epoch ($z=1100$) and has no influence on the estimation of $\delta_m$ at $z=1100$. We solved the differential equations numerically from $z=0$ to $z=1100$ and found that $\delta_m$ is increasing in the positive direction, whereas $\delta_{de}$ is increasing in the negative direction. However, as $\delta_m$ and $\delta_{de}$ are always scaled by their respective present values, this difference in signature is not reflected in the plots. 

A small negative value of $\beta$ is found to be indicated for the density contrast to grow. This suggests that the dark matter decays into dark energy and the interaction in the dark sector, if any, has to be small. 

The effect of effective sound speed of DE, $c_{s,de}^2$ on density perturbation was also looked at. The absence of an interaction, with $w_{de} =-1$ and $c_{s,de}^2=0$ indicates no pressure perturbation in DE and the DE density perturbation is expected to grow like that of DM. In the present HDE model, the scenario is entirely different; even in the absence of interaction and $c_{s,de}^2$, the pressure perturbation remains non-vanishing as $w_{de}$ is now a varying function of redshift, $z$. The pressure perturbation is then governed by the velocity perturbation through $\theta_{de}$ and the background quantities $\rho_{de}$, $w_{de}$ and $Q_{de}$.

The DE density contrast, $\delta_{de}$ also grows almost in a similar fashion like its DM counterpart, $\delta_m$, for most of the evolution after the radiation dominated era, but right at the present moment is actually decaying after hitting a maximum in the recent past. This is true even in the absence of interaction (characterised by $\beta=0$). This maximum is found to occur after the onset of the present accelerated phase of expansion. The height of the maximum is related to the strength of the interaction, $\beta$. On decreasing the strength (smaller magnitude) of the interaction, the position of maximum shifts to lower redshifts and the height decreases. This feature is observed for the zero value of the effective sound speed, $c_{s,de}^2$(figure \ref{figcs} b).

When $c_{s,de}^2=0$ the first part in the expression for $\delta p_{de}$ (see equation (\ref{pert-p})) is zero and from the second part, we can say that $\delta_{de}$ reaching a maximum is characterised by $\theta_{de}$ whereas $Q_{de}$ actually suppresses this feature (figures \ref{figmul0183} b and \ref{figmul0175} b). When $c_{s,de}^2 \ne 0$, the contribution from the first part ($c_{s,de}^2 \delta \rho_{de}$) results in the steep rise in $\delta_{de}$ at lower values of $z$ (figure \ref{figcs} b). This apparently is engineered by $\theta_{de}$. Except for the peak in the growth rate, similar features are also seen for $\delta_{m}$ (figure \ref{figcs} a).
For $c_{s,de}^2=1$, a rapid short-lived oscillations in the DE density contrast is found between $z=0.1275$ to $z=0.4875$. These oscillations are characteristic of $c_{s,de}^2=1$ and $\beta \ne 0$ (figure \ref{figcs1}). 

For an interaction in dark sector, without a holographic bound, there is an instability in the perturbation\cite{valiviita2008jcap}. This does not appear in the present case where there is a holographic bound.

Thus the present investigation has some new inputs leading to quite new and intriguing features. The new physical input at the outset is certainly the introduction of an interaction between DM and DE in the study of density contrasts. In the techniques and approximations, writing down the full relativistic perturbation equations a priori is new. Even in the most recent and general treatment\cite{mehrabi}, $\Phi^\prime$ is neglected for the estimations, but in the present work, its contribution is also respected. The appearance of a peak $\delta_{de}$ for $c^{2}_{s,de} =0$ is a completely new feature observed, which is not due to the interaction, as it is there even for $\beta =0$~. So this is due to the inclusion of $\Phi^\prime$ in the estimation. For $c^{2}_{s,de} =1$, no growth for $\delta_{de}$ is normally observed. The recent work of Batista and Pace\cite{batista2013} shows an almost negligible growth. The present work shows a very steep growth for small $z$. This is there both in the presence and absence of interaction. For an interacting scenario, there is also a short-lived oscillatory period in the growing mode of $\delta_{de}$ (figures \ref{figcs1} a and \ref{figcs1} b).

Although we presented the calculations with $\phi_0=10^{-5}$ and $k=0.1h$, these results are insensitive to changes in $\phi_0$ and $k$ mode entering the horizon. 

\begin{appendices}
\section{Coefficients of the coupled differential equations} \label{appendix}

The coefficients of equations (\ref{finaldm}) and (\ref{finalde}) are given below.
\begin{itemize}
\item[(i)] The coefficients of equation (\ref{finaldm}) are:
\begin{equation}
\mathbf{C^{(m)}_1} = -H_0 E  ~,
\end{equation}
\begin{equation}
\mathbf{C^{(m)}_2} =
-\frac{E   ((2 \beta -1) H_0 \Omega_{de}  +H_0)}{(z+1) (\Omega_{de}  -1)}-H_0 E ' ~,
\end{equation}
\begin{equation}
\begin{split}
\mathbf{C^{(m)}_3} =
\frac{-\left(H_0 \beta  \left((z+1) (\Omega_{de}  -1) \Omega_{de}   E ' + E \left(2 \Omega_{de}  +(\beta -2) \Omega_{de}^2- (z+1) \Omega_{de} ' \right)\right)\right)}{(z+1)^2 (\Omega_{de}  -1)^2} ~,
\end{split}
\end{equation}
\begin{equation}
\mathbf{C^{(m)}_4} = 0 ~,
\end{equation}
\begin{equation}
\begin{split}
\mathbf{C^{(m)}_5} = 
& \left(H_0 \beta  E   \Omega_{de}   \left(-9 H_0^2 w_{de}^2 E^2+\right.\right.
 w_{de}   \left(3 H_0^2 \left(-\beta +3 c_{s,de}^2-3\right) E^2+k^2 (z+1)^2\right)+\\
& \left.\left.\left.3 H_0^2 E^2 \left((\beta +3) c_{s,de}^2-(z+1) w_{de} ' \right)\right)\right)\right/(z+1 
 \Omega_{de}  -1 \left(-k^2 (z+1)^2-9 H_0^2 w_{de}^2 E^2+\right.\\
& w_{de}   \left(3 H_0^2 \left(-\beta +3 c_{s,de}^2-3\right) E^2+k^2 (z+1)^2\right)+
 \left.\left.3 H_0^2 E^2 \left((\beta +3) c_{s,de}^2-(z+1) w_{de} ' \right)\right)\right) ~,
\end{split}
\end{equation}
\begin{equation}
\begin{split}
\mathbf{C^{(m)}_6} = 
& (H_0 \beta  (E   \Omega_{de}  -
 E   (\Omega_{de}  -1) \Omega_{de}  -
 E   (\Omega_{de}^2+\beta  E   (\Omega_{de}^2+
 \left.3 c_{s,de}^2 k^2 (z+1) E   (\Omega_{de}  -1) \Omega_{de}  \right/
 \left(k^2 (z+1)^2+\right.\\
& \left.9 H_0^2 w_{de}^2 E^2-\right.
 w_{de}   \left(3 H_0^2 \left(-\beta +3 c_{s,de}^2-3\right) E^2+k^2 (z+1)^2\right)+\\
& \left.3 H_0^2 E^2 \left((z+1) w_{de} ' -(\beta +3) c_{s,de}^2\right)\right)+
 \left.3 c_{s,de}^2 k^2 z (z+1) E   (\Omega_{de}  -1) \Omega_{de}  \right/
 \left(k^2 (z+1)^2+9 H_0^2 w_{de}^2 E^2-\right.\\
& w_{de}   \left(3 H_0^2 \left(-\beta +3 c_{s,de}^2-3\right) E^2+k^2 (z+1)^2\right)+\\
& \left.3 H_0^2 E^2 \left((z+1) w_{de} ' -(\beta +3) c_{s,de}^2\right)\right)-
 \left(3 k^2 (z+1) w_{de}   E  \right.
 (\Omega_{de}  -1) \Omega_{de}  )/\\
& \left(k^2 (z+1)^2+9 H_0^2 w_{de}^2 E^2-\right.
 w_{de}   \left(3 H_0^2 \left(-\beta +3 c_{s,de}^2-3\right) E^2+k^2 (z+1)^2\right)+\\
& \left.3 H_0^2 E^2 \left((z+1) w_{de} ' -(\beta +3) c_{s,de}^2\right)\right)-
 \left(3 k^2 z (z+1) w_{de}   E  \right.\\
& (\Omega_{de}  -1) \Omega_{de}  )/
 \left(k^2 (z+1)^2+9 H_0^2 w_{de}^2 E^2-\right.
 w_{de}   \left(3 H_0^2 \left(-\beta +3 c_{s,de}^2-3\right) E^2+k^2 (z+1)^2\right)+\\
& \left.3 H_0^2 E^2 \left((z+1) w_{de} ' -(\beta +3) c_{s,de}^2\right)\right)+
 (z+1) (\Omega_{de}  -1) \Omega_{de}   E ' +\\
& (z+1) E   (\Omega_{de}  -1) \Omega_{de} ' -
 \left.\left.\left.(z+1) E   \Omega_{de}   \Omega_{de} ' \right)\right)\right/
 (z+1)^2 (\Omega_{de}  -1)^2 ~,
\end{split}
\end{equation}
\begin{equation}
\mathbf{C^{(m)}_7} = 3 H_0 E   ~,
\end{equation}
\begin{equation}
\begin{split}
\mathbf{C^{(m)}_8} = 
& \left(H_0 \left(3 \beta  k^2 E   \Omega_{de}  +6 \beta  k^2 z E  \right.\right.
 \Omega_{de}  +3 \beta  k^2 z^2 E   \Omega_{de}  -
 3 \beta  k^2 w_{de}   E   \Omega_{de}  -
 6 \beta  k^2 z w_{de}   E   \Omega_{de}  -
 3 \beta  k^2 z^2 w_{de}   E   \Omega_{de}  -\\
& 3 E   \left(k^2 (z+1)^2+9 H_0^2 w_{de}^2 E^2-\right.
 w_{de}   \left(3 H_0^2 \left(-\beta +3 c_{s,de}^2-3\right) E^2+k^2 (z+1)^2\right)+\\
& \left.3 H_0^2 E^2 \left((z+1) w_{de} ' -(\beta +3) c_{s,de}^2\right)\right)+
 3 E   \Omega_{de}  
 \left(k^2 (z+1)^2+9 H_0^2 w_{de}^2 E^2-\right.\\
& w_{de}   \left(3 H_0^2 \left(-\beta +3 c_{s,de}^2-3\right) E^2+k^2 (z+1)^2\right)+
 \left.3 H_0^2 E^2 \left((z+1) w_{de} ' -(\beta +3) c_{s,de}^2\right)\right)-
 4 \beta  E   \Omega_{de}  \\
& \left(k^2 (z+1)^2+9 H_0^2 w_{de}^2 E^2-\right.
 w_{de}   \left(3 H_0^2 \left(-\beta +3 c_{s,de}^2-3\right) E^2+k^2 (z+1)^2\right)+\\
& \left.3 H_0^2 E^2 \left((z+1) w_{de} ' -(\beta +3) c_{s,de}^2\right)\right)+
 3 (z+1) (1-\Omega_{de}  )
 \left(k^2 (z+1)^2+9 H_0^2 w_{de}^2 E^2-\right.\\
& w_{de}   \left(3 H_0^2 \left(-\beta +3 c_{s,de}^2-3\right) E^2+k^2 (z+1)^2\right)+
 \left.3 H_0^2 E^2 \left((z+1) w_{de} ' -(\beta +3) c_{s,de}^2\right)\right) \\
& \left.\left.\left.E ' \right)\right)\right/((z+1) (1-\Omega_{de}  )
 \left(k^2 (z+1)^2+9 H_0^2 w_{de}^2 E^2-\right.
 w_{de}   \left(3 H_0^2 \left(-\beta +3 c_{s,de}^2-3\right) E^2+k^2 (z+1)^2\right)+\\
& \left.\left.3 H_0^2 E^2 \left((z+1) w_{de} ' -(\beta +3) c_{s,de}^2\right)\right)\right) ~,
\end{split}
\end{equation}
\begin{equation}
\begin{split}
\mathbf{C^{(m)}_9} = 
& -\frac{k^2}{H_0 E  }+\frac{\beta  H_0 \Omega_{de}   E ' }{(z+1) (\Omega_{de}  -1)}+
 (H_0 \beta  E   (\Omega_{de}  -(\Omega_{de}  -1)
 \Omega_{de}  -(\Omega_{de}^2+\beta  (\Omega_{de}^2+
 \left.\beta  k^2 (z+1) (\Omega_{de}  -1) \Omega_{de}  \right/\\
& \left(k^2 (z+1)^2+9 H_0^2 w_{de}^2 E^2-\right.
 w_{de}   \left(3 H_0^2 \left(-\beta +3 c_{s,de}^2-3\right) E^2+k^2 (z+1)^2\right)+\\
& \left.3 H_0^2 E^2 \left((z+1) w_{de} ' -(\beta +3) c_{s,de}^2\right)\right)+
 \left.\beta  k^2 z (z+1) (\Omega_{de}  -1) \Omega_{de}  \right/
 \left(k^2 (z+1)^2+9 H_0^2 w_{de}^2 E^2-\right.\\
& w_{de}   \left(3 H_0^2 \left(-\beta +3 c_{s,de}^2-3\right) E^2+k^2 (z+1)^2\right)+
 \left.3 H_0^2 E^2 \left((z+1) w_{de} ' -(\beta +3) c_{s,de}^2\right)\right)+
 (z+1) (\Omega_{de}  -1) \Omega_{de} ' -\\
& \left.\left.\left.(z+1) \Omega_{de}   \Omega_{de} ' \right)\right)\right/
 (z+1)^2 (\Omega_{de}  -1)^2 ~.
\end{split}
\end{equation}
\item[(ii)] The coefficients of equation (\ref{finalde}) are:
\begin{equation}
\mathbf{C^{(de)}_1}=-H_0 E  ~,
\end{equation}
\begin{equation}
\begin{split}
\mathbf{C^{(de)}_2}=
& -\left(\left(H_0 \left(-k^2 (z+1)^2 E   \left(-1-\beta  c_{s,de}^2+\right.\right.\right.\right.
 (\Omega_{de}^2-3 (\Omega_{de}^3+(z+1) w_{de} ' +\\
& \left.w_{de}   \left((z+1) w_{de} ' +\beta  c_{s,de}^2+3\right)\right)+
 k^2 (z+1)^3 \left((\Omega_{de}^2-1\right) E ' +\\
& 3 H_0^2 (z+1) (w_{de}  +1) E^2
  \left((\beta +3) \left(-c_{s,de}^2\right)+\left(\beta -3 c_{s,de}^2+3\right) w_{de}  +\right.\\
& \left.(z+1) w_{de} ' +3 (\Omega_{de}^2\right) E ' +
 3 H_0^2 E^3 \left(3 c_{s,de}^2+\beta  c_{s,de}^2-3 \beta  c_{s,de}^4-\beta ^2 c_{s,de}^4+\right.\\
& 3 \left(-\beta +3 c_{s,de}^2-7\right) (\Omega_{de}^3-9 (\Omega_{de}^4+
 (z+1) \left(\beta +(\beta -3) c_{s,de}^2+3\right) w_{de} ' +(\Omega_{de}^2\\
&  3 (z+1) w_{de} ' -4 \beta +3 (2 \beta +7) c_{s,de}^2-15+
(z+1)^2 w_{de} ''  +\\
& w_{de}   \left(-\beta -3 \beta  c_{s,de}^4+\beta ^2 c_{s,de}^2+7 \beta  c_{s,de}^2+15 c_{s,de}^2-(z+1)-3\right.\\
& \left.\left.\left.\left.\left. \left(-\beta +3 c_{s,de}^2-6\right) w_{de} ' +(z+1)^2 w_{de} '' \right)\right)\right)\right)\right/
 \left(z+1 w_{de}  +1 \left(-9 H_0^2 (\Omega_{de}^2-k^2 (z+1)^2\right.\right.
 E^2+\\
& w_{de}   \left(3 H_0^2 \left(-\beta +3 c_{s,de}^2-3\right) E^2+k^2 (z+1)^2\right)+
 \left.\left.\left.3 H_0^2 E^2 \left((\beta +3) c_{s,de}^2-(z+1) w_{de} ' \right)\right)\right)\right) ~,
\end{split}
\end{equation}
\begin{equation}
\begin{split}
\mathbf{C^{(de)}_3}=
& \left(c_{s,de}^2 k^4 (z+1)^4-3 H_0^2 k^2 (z+1)^2 E^2\right.
 -\left(3 c_{s,de}^2-1\right) (z+1) w_{de} ' +3 (\beta +1) c_{s,de}^4+2 c_{s,de}^2+\\
& 3 c_{s,de}^2 H_0^2 k^2 (z+1)^3 E   E ' +
 27 H_0^4 (z+1) (\Omega_{de}^4 E^3 E ' +
 9 c_{s,de}^2 H_0^4 (z+1) E^3 \left((\beta +3) c_{s,de}^2-(z+1) w_{de} ' \right)\\
& 3 H_0^2 (\Omega_{de}^3 E  +E ' 
 \left(9 H_0^2 E^3-\left(3 c_{s,de}^2+2\right) k^2 (z+1)^2 E  \right.
 (z+1) w_{de} ' +2 \beta  c_{s,de}^2+k^2 (z+1)^3 E ' -\\
& \left.3 H_0^2 (z+1) \left(-\beta +6 c_{s,de}^2-6\right) E^2 E ' \right)+
 9 H_0^4 E^4 \left(2 \beta  (\beta +3) c_{s,de}^6-c_{s,de}^2 (z+1) \left(3 \beta  c_{s,de}^2+1\right)\right.\\
&  w_{de} ' +\left(c_{s,de}^2-1\right) (z+1)^2 \left(w_{de} ' \right)^2-
 \left.c_{s,de}^2 (z+1)^2 w_{de} '' \right)+\\
&(\Omega_{de}^2 
 \left(c_{s,de}^2 k^4 (z+1)^4+3 c_{s,de}^2 H_0^2 k^2 (z+1)^2 \left(-3 \beta +3 c_{s,de}^2+2\right) E^2-\right.\\
& 3 c_{s,de}^2 H_0^2 k^2 (z+1)^3 E   E ' +9 H_0^4 (z+1)
 E^3 \left((z+1) w_{de} ' +\beta +3 c_{s,de}^4-2 (\beta +6) c_{s,de}^2+3\right)\\
&  E ' -9 H_0^4 E^4 \left(2 \beta  c_{s,de}^2 \left(-\beta +6 c_{s,de}^2-3\right)+\right.
 \left.\left.\left(3 c_{s,de}^2-4\right) (z+1) w_{de} ' -(z+1)^2 w_{de} '' \right)\right)+\\
& w_{de}   \left(-3 H_0^2 k^2 (z+1)^2 E^2-2 c_{s,de}^2 k^4 (z+1)^4\right.\\
& \left(\left(c_{s,de}^2+1\right) (z+1) w_{de} ' -3 \beta  c_{s,de}^4-3 (\beta +1) c_{s,de}^2-2\right)-3 H_0^2
 9 H_0^4 (z+1) E^3+k^2 (z+1)^3 E   E ' \\
& -\left(c_{s,de}^2-1\right) (z+1) w_{de} ' +(\beta +6) c_{s,de}^4-2 (\beta +3) c_{s,de}^2
 9 H_0^4 E^4+E ' \\
& \left(6 \beta  c_{s,de}^6-4 \beta  (\beta +3) c_{s,de}^4+(z+1) \left((3 \beta -4) c_{s,de}^2+1\right) w_{de} ' -\right.\\
& \left.\left.\left.\left.(z+1)^2 \left(w_{de} ' \right)^2-\left(c_{s,de}^2-1\right) (z+1)^2 w_{de} '' \right)\right)\right)\right/
 \left(H_0 (z+1)^2 (w_{de}  +1) E  \right.
 \left(k^2 (z+1)^2+9 H_0^2 (\Omega_{de}^2 E^2-\right.\\
& w_{de}   \left(3 H_0^2 \left(-\beta +3 c_{s,de}^2-3\right) E^2+k^2 (z+1)^2\right)+
 \left.\left.3 H_0^2 E^2 \left((z+1) w_{de} ' -(\beta +3) c_{s,de}^2\right)\right)\right) ~,
\end{split}
\end{equation}
\begin{equation}
\begin{split}
\mathbf{C^{(de)}_4}= \mathbf{C^{(de)}_5}=\mathbf{C^{(de)}_6}=0 ~,
\end{split}
\end{equation}
\begin{equation}
\begin{split}
\mathbf{C^{(de)}_7}=3 H_0 (w_{de}  -1) E~,
\end{split}
\end{equation}
\begin{equation}
\begin{split}
\mathbf{C^{(de)}_8}=
& \left(H_0 k^2 (z+1)^2 (w_{de}  -1)\right.
 \left(3+3 (\beta -3) c_{s,de}^2+\beta +\left(\beta -3 \beta  c_{s,de}^2\right) w_{de}  +\right.\\
& \left.\left(9 c_{s,de}^2-3\right) (\Omega_{de}^2\right) E  +
 3 H_0 k^2 (z+1)^3 (w_{de}  -1)^2 (w_{de}  +1) E ' +\\
& 9 H_0^3 (z+1) \left((\Omega_{de}^2-1\right) E^2
 \left((\beta +3) \left(-c_{s,de}^2\right)+\left(\beta -3 c_{s,de}^2+3\right) w_{de}  +3 (\Omega_{de}^2+\right.\\
& \left.(z+1) w_{de} ' \right) E ' -3 H_0^3 E^3
 \left(9 c_{s,de}^2+27 c_{s,de}^4-\beta ^2 c_{s,de}^2-3 \beta ^2 c_{s,de}^4+9 \left(3 c_{s,de}^2+1\right) (\Omega_{de}^4-\right.\\
& (z+1) \left(-4 \beta +27 c_{s,de}^2-9\right) w_{de} ' +3 (z+1)^2 \left(w_{de} ' \right)^2-\\
& 3 (\Omega_{de}^3 \left(3 (z+1) w_{de} ' -2 \beta +9 c_{s,de}^4-6 c_{s,de}^2-3\right)+\\
& w_{de}   \left(-9-18 c_{s,de}^2+27 c_{s,de}^4-6 \beta  c_{s,de}^2+\beta ^2+2 \beta ^2 c_{s,de}^2+\right.\\
& 3 \beta ^2 c_{s,de}^4-(z+1) \left(-4 \beta +6 (\beta +3) c_{s,de}^2-27\right) w_{de} ' +
 \left.3 (z+1)^2 \left(w_{de} ' \right)^2\right)+\\
& 3 w_{de} '' +6 z w_{de} '' +3 z^2 w_{de} '' +
 (\Omega_{de}^2 \left(-9-36 c_{s,de}^2-27 c_{s,de}^4+6 \beta -6 \beta  c_{s,de}^2+\beta ^2-3 \beta ^2 c_{s,de}^2+\right.\\
& \left.\left.\left.\left.9 \left(c_{s,de}^2+1\right) (z+1) w_{de} ' -3 (z+1)^2 w_{de} '' \right)\right)\right)\right/
 \left(z+1 w_{de}  +1 \left(-k^2 (z+1)^2-\right.\right.\\
& 9 H_0^2 (\Omega_{de}^2 E^2+
 w_{de}   \left(3 H_0^2 \left(-\beta +3 c_{s,de}^2-3\right) E^2+k^2 (z+1)^2\right)+
 \left.\left.3 H_0^2 E^2 \left((\beta +3) c_{s,de}^2-(z+1) w_{de} ' \right)\right)\right) ~,
\end{split}
\end{equation}
\begin{equation}
\begin{split}
\mathbf{C^{(de)}_9}=
& \left(81 H_0^4 (\Omega_{de}^5 E^4-9 (\Omega_{de}^4+k^4 (z+1)^4\right.
 3 H_0^4 \left(-2 \beta +6 c_{s,de}^2-9\right) E^4+2 H_0^2 k^2 (z+1)^2 E^2+\\
& H_0^2 k^2 (z+1)^2 E^2
 (\beta +6) (z+1) w_{de} ' -2 \beta -\left(\beta ^2+3 \beta +18\right) c_{s,de}^2+
 \beta  H_0^2 k^2 (z+1)^3 E   E ' -\\
& 3 \beta  H_0^4 (z+1) E^3 \left((z+1) w_{de} ' -(\beta +3) c_{s,de}^2\right)
 E ' +(\Omega_{de}^3 \left(k^4 (z+1)^4+\right.\\
& H_0^2 k^2 (z+1)^2 \left(-\beta +3 c_{s,de}^2-3\right) E^2+9 H_0^4 E^4\\
& 6 (z+1) w_{de} ' +\beta ^2+12 \beta +9 c_{s,de}^4-9 (\beta +6) c_{s,de}^2+27-
 \left.9 \beta  H_0^4 (z+1) E^3 E ' \right)-\\
& (\Omega_{de}^2 \left(H_0^2 k^2 (z+1)^2 E^2+k^4 (z+1)^4\right.
  6 (z+1) w_{de} ' -2 (\beta +9)-3 (\beta +6) c_{s,de}^2-\\
& 9 H_0^4 E^4 \left((\beta +3)^2+3 (\beta +9) c_{s,de}^4-2 \left(\beta ^2+9 \beta +27\right) c_{s,de}^2-\right.
 \left.6 \left(c_{s,de}^2-2\right) (z+1) w_{de} ' \right)+\\
& \beta  H_0^2 k^2 (z+1)^3 E   E ' -
 \left.3 \beta  H_0^4 (z+1) \left(-\beta +3 c_{s,de}^2-6\right) E^3 E ' \right)+\\
& 3 H_0^4 E^4 \left(\left(\beta ^3+3 \beta ^2+9 \beta +27\right) c_{s,de}^4-\right.
 (z+1) \left(\beta  (\beta +4)+\left(\beta ^2+18\right) c_{s,de}^2\right) w_{de} ' +\\
& \left.3 (z+1)^2 \left(w_{de} ' \right)^2-\beta  (z+1)^2 w_{de} '' \right)-
 w_{de}   \left(k^4 (z+1)^4-H_0^2 k^2 (z+1)^2 E^2\right.\\
& \beta  (z+1) w_{de} ' +6 (\beta +3)+\left(\beta ^2-18\right) c_{s,de}^2+
 3 \beta  H_0^4 (z+1) E^3\\
& E '  \left((z+1) w_{de} ' +\beta -(\beta +6) c_{s,de}^2+3\right)-\\
& 3 H_0^4 E^4 \left(c_{s,de}^2 \left(-\beta ^3-6 \beta ^2-27 \beta +3 \left(\beta ^2+6 \beta +27\right) c_{s,de}^2-54\right)-\right.\\
& (z+1) \left(\beta ^2+4 \beta +36 c_{s,de}^2-18\right) w_{de} ' +
 \left.\left.\left.\left.3 (z+1)^2 \left(w_{de} ' \right)^2-\beta  (z+1)^2 w_{de} '' \right)\right)\right)\right/\\
& \left(H_0 (z+1)^2 (w_{de}  +1) E  \right.
 \left(k^2 (z+1)^2+9 H_0^2 (\Omega_{de}^2 E^2-\right.
 w_{de}   \left(3 H_0^2 \left(-\beta +3 c_{s,de}^2-3\right) E^2+k^2 (z+1)^2\right)+\\
& \left.\left.3 H_0^2 E^2 \left((z+1) w_{de} ' -(\beta +3) c_{s,de}^2\right)\right)\right) ~.
\end{split}
\end{equation}
\end{itemize}
\end{appendices}

\end{document}